\documentclass{article}
\pdfoutput=1

\usepackage{arxiv}
\usepackage[utf8]{inputenc} % allow utf-8 input
\usepackage[T1]{fontenc} % use 8-bit T1 fonts
\usepackage{hyperref} % hyperlinks
\hypersetup{
	colorlinks = true,
	linkcolor = black,
	anchorcolor = black,
	citecolor = black,
	filecolor = black,
	urlcolor = blue
}
\usepackage{booktabs} % professional-quality tables
\usepackage{amsfonts} % blackboard math symbols
\usepackage{nicefrac} % compact symbols for 1/2, etc.
\usepackage{microtype} % microtypography
\usepackage{amsmath,bm}
\usepackage{graphicx}
\usepackage[flushleft]{threeparttable}
\usepackage{multirow}
\usepackage{natbib}
\usepackage{makecell}
\usepackage{caption}
\usepackage{subcaption}   
\usepackage{float}
\usepackage{nicematrix}

\usepackage{array}
\newcommand{\PreserveBackslash}[1]{\let\temp=\\#1\let\\=\temp}
\newcolumntype{C}[1]{>{\PreserveBackslash\centering}p{#1}}
\newcolumntype{R}[1]{>{\PreserveBackslash\raggedleft}p{#1}}
\newcolumntype{L}[1]{>{\PreserveBackslash\raggedright}p{#1}}

\title{Jenss-Bayley Latent Change Score Model with Individual Ratio of Growth Acceleration in the Framework of Individual Measurement Occasions}

\author{
  Jin Liu \thanks{CONTACT Jin Liu Email: Veronica.Liu0206@gmail.com, \textcircled{c}2022, Journal of Educational and Behavioral Statistics. This paper is not the copy of record and may not exactly replicate the final, authoritative version of the article. Please do not copy or cite without authors' permission. }\\
  Department of Biostatistics\\
  Virginia Commonwealth University \\
}

\begin{document}
\maketitle
\begin{abstract}
Longitudinal analysis has been widely employed to examine between-individual differences in within-individual change. One challenge of such analyses lies in that the rate-of-change is only available indirectly when change patterns are nonlinear with respect to time. Latent change score models (LCSMs), which can be employed to investigate the change in growth rate at the individual level, have been developed to address this challenge. We extend an existing LCSM with the Jenss-Bayley growth curve \citep[Chapter~18]{Grimm2016growth} and propose a novel expression of change scores that allows for (1) unequally-spaced study waves and (2) individual measurement occasions around each wave. We also extend the existing model to estimate the individual ratio of growth acceleration (that largely determines the trajectory shape and is viewed as the most important parameter in the Jenss-Bayley model). We present the proposed model by simulation studies and a real-world data analysis. Our simulation studies demonstrate that the proposed model generally estimates the parameters of interest unbiasedly, precisely, and exhibits appropriate confidence interval coverage. More importantly, the proposed model with the novel expression of change scores performed better than the existing model shown by simulation studies. An empirical example using longitudinal reading scores shows that the model can estimate the individual ratio of growth acceleration and generate individual growth rate in practice. We also provide the corresponding code of the proposed model. 
\end{abstract}


\section{Introduction}\label{sec:Intro}
Researchers use latent growth curve models to examine within-individual changes and between-individual differences simultaneously. One coefficient out of the most research interest is the rate-of-change, which can only be directly estimated in a linear model. However, if the study duration is long enough, the change patterns show a nonlinear relationship with time $t$. Accordingly, empirical researchers often assume that the trajectories take nonlinear parametric functional forms, such as quadratic, exponential, and Jenss-Bayley functions. In these models, the rate-of-change does not appear explicitly, and therefore, the between-individual differences in the rate-of-change cannot be analyzed directly.

Fortunately, multiple remedies have been proposed to address this challenge. For example, \citet{Harring2006nonlinear, Kohli2015PLGC1, Liu2021PBLSGM, Harring2021piece} recommended utilizing piecewise functional forms, such as bilinear spline (i.e., linear-linear piecewise) or more linear pieces, where the mean and variance of the rate-of-change of each segment can be estimated directly to capture the underlying change patterns. One challenge of these semi-parametric functions is that researchers have to decide the transition time from one linear piece to another. The detailed discussion of the transition time can be found in earlier studies, such as \citet{Kohli2015PLGC1, Liu2021PBLSGM}. Alternatively, \citet{Grimm2013LCM1}, \citet{Grimm2013LCM2}, and \citet[Chapter~18]{Grimm2016growth} have demonstrated how to implement a latent change score model (LCSM), which can be viewed as the first derivative of the corresponding latent growth curve model (LGCM) with respect to time $t$, to investigate the instantaneous rate-of-change. This present study proposes a novel specification for the LCSM with parametric functional forms to allow (1) unequally-spaced study waves and (2) individual measurement occasions around each wave. Specifically, we demonstrate how to apply this specification to the LCSM with the Jenss-Bayley growth curve. 
\subsection{Introduction of Jenss-Bayley Function}\label{I:JB}
\begin{figure}[!htbp]
\centering
\begin{subfigure}{.5\textwidth}
  \centering
  \includegraphics[width=1.0\linewidth]{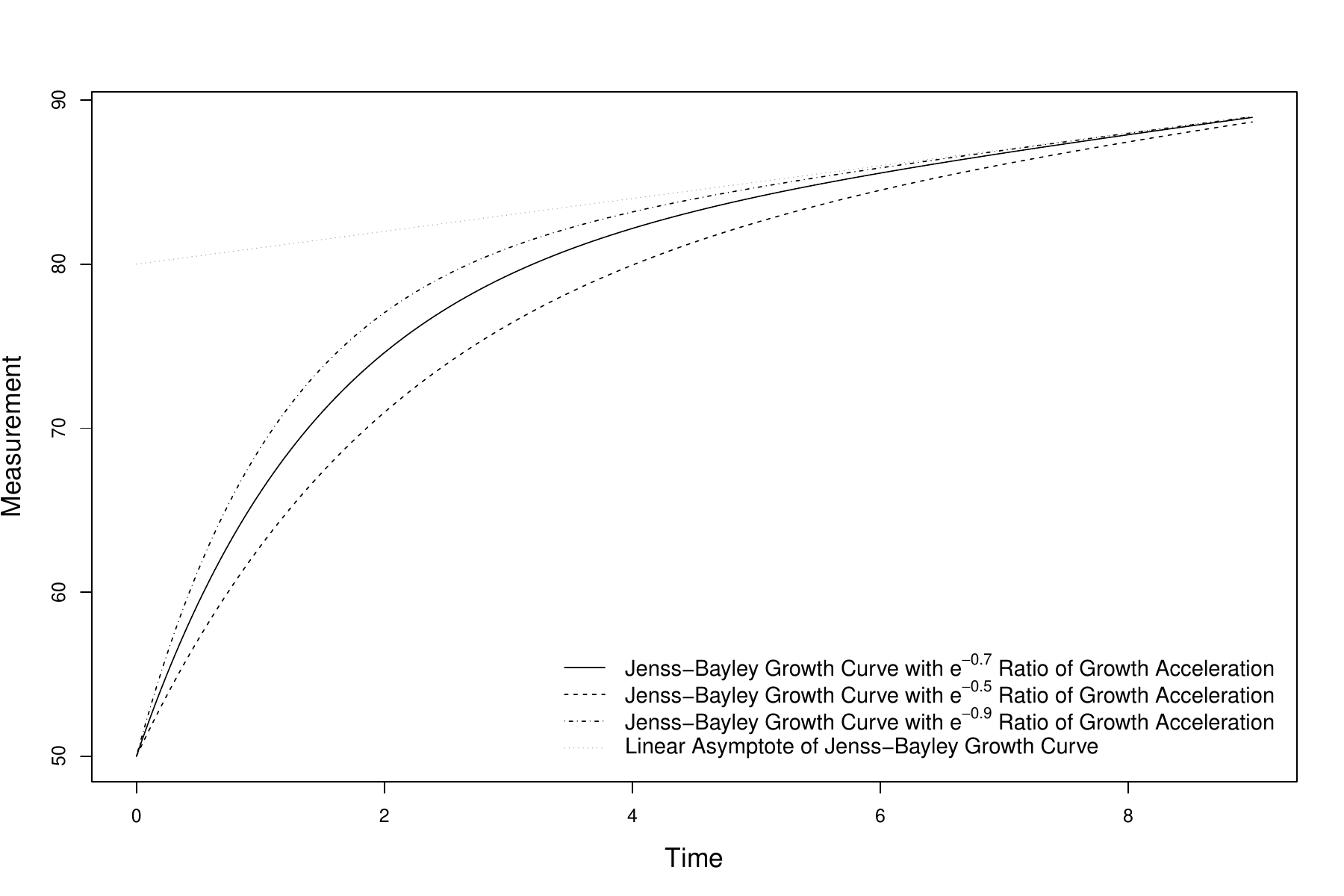}
  \caption{Growth Curve}
  \label{fig:JB_LGC}
\end{subfigure}%
\begin{subfigure}{.5\textwidth}
  \centering
  \includegraphics[width=1.0\linewidth]{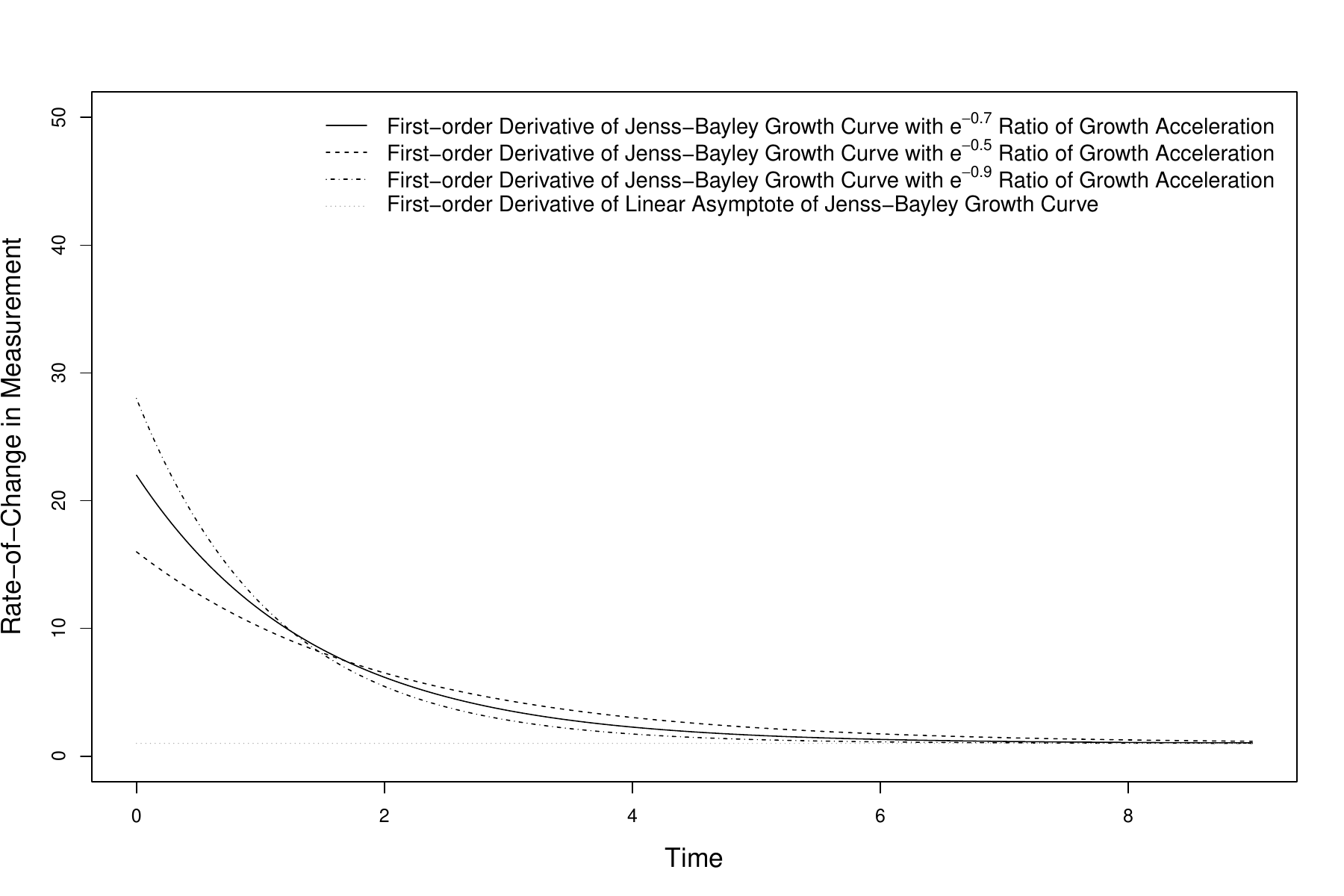}
  \caption{Instantaneous Rate-of-Change over Time}
  \label{fig:JB_LCM}
\end{subfigure}
\caption{Jenss-Bayley Trajectory and Its Instantaneous Rate-of-Change with Different Ratio of the growth acceleration (Values of Other Coefficients: $\eta_{0}=50$; $\eta_{1}=1.0$; $\eta_{2}=-30$)}
\label{fig:JB}
\end{figure}

The Jenss-Bayley model is a four-parameter nonlinear model described by \citet{Jenss1937JB}, which can be viewed as a combination of linear and exponential growth models. Its functional form is as follows
\begin{equation}\nonumber
y_{j}=a_{0}+a_{1}t_{j}-\exp(c_{0}+c_{1}t_{j})+\epsilon,
\end{equation}
which is a negative-accelerated exponential that approaches a linear asymptote with a positive slope. In the function, $y_{j}$ and $t_{j}$ are the measurement and measurement occasion at time $j$, $a_{0}$ and $a_{1}$ are the intercept and slope of the linear asymptote, respectively, $\exp(c_{0})$ is the vertical distance between the initial status and the intercept of the linear asymptote, and $\exp(c_{1})$ is the growth or acceleration constant that measures the ratio of acceleration of growth\footnote{By `the acceleration of growth', we mean the second derivative of the growth curve with respect to time $t$, or equivalently, the first derivative of the rate-of-change with respect to time $t$. } at time $j$ to that at the preceding time $j-1$ \citep{Jenss1937JB}. According to  \citet[Chapter~11]{Grimm2016growth}, we can write the model as 
\begin{equation}\nonumber
y_{j}=\eta_{0}+\eta_{1}t_{j}+\eta_{2}(\exp(\gamma t_{j})-1)+\epsilon,
\end{equation}
where $\eta_{0}=a_{0}-\exp(c_{0})$ (i.e., $\eta_{0}$ is the initial status), $\eta_{1}=a_{1}$, $\eta_{2}=-\exp(c_{0})$ (i.e., $\eta_{2}$ is the vertical distance between two intercepts), and $\gamma=c_{1}$. According to \citet{Jenss1937JB}, the magnitude of $\exp(\gamma)$ is what largely determines the trajectory shape (see Figures \ref{fig:JB_LGC} and \ref{fig:JB_LCM} for the Jenss-Bayley growth curve and growth rate with three different $\gamma$'s: $\gamma=-0.9, -0.7, -0.5$). From the figures, we can see that a more negative $\gamma$ leads to a more curvature at the early stage, with a steep initial development followed by level-off growth and an earlier approach to the linear asymptote. In contrast, a less negative $\gamma$ results in a flatter curve with a later approach to the asymptote.

\subsection{Introduction of Latent Change Score Models}\label{I:LCSM}
\begin{figure}[!htbp]
\centering
\begin{subfigure}{.5\textwidth}
  \centering
  \includegraphics[width=0.95\linewidth]{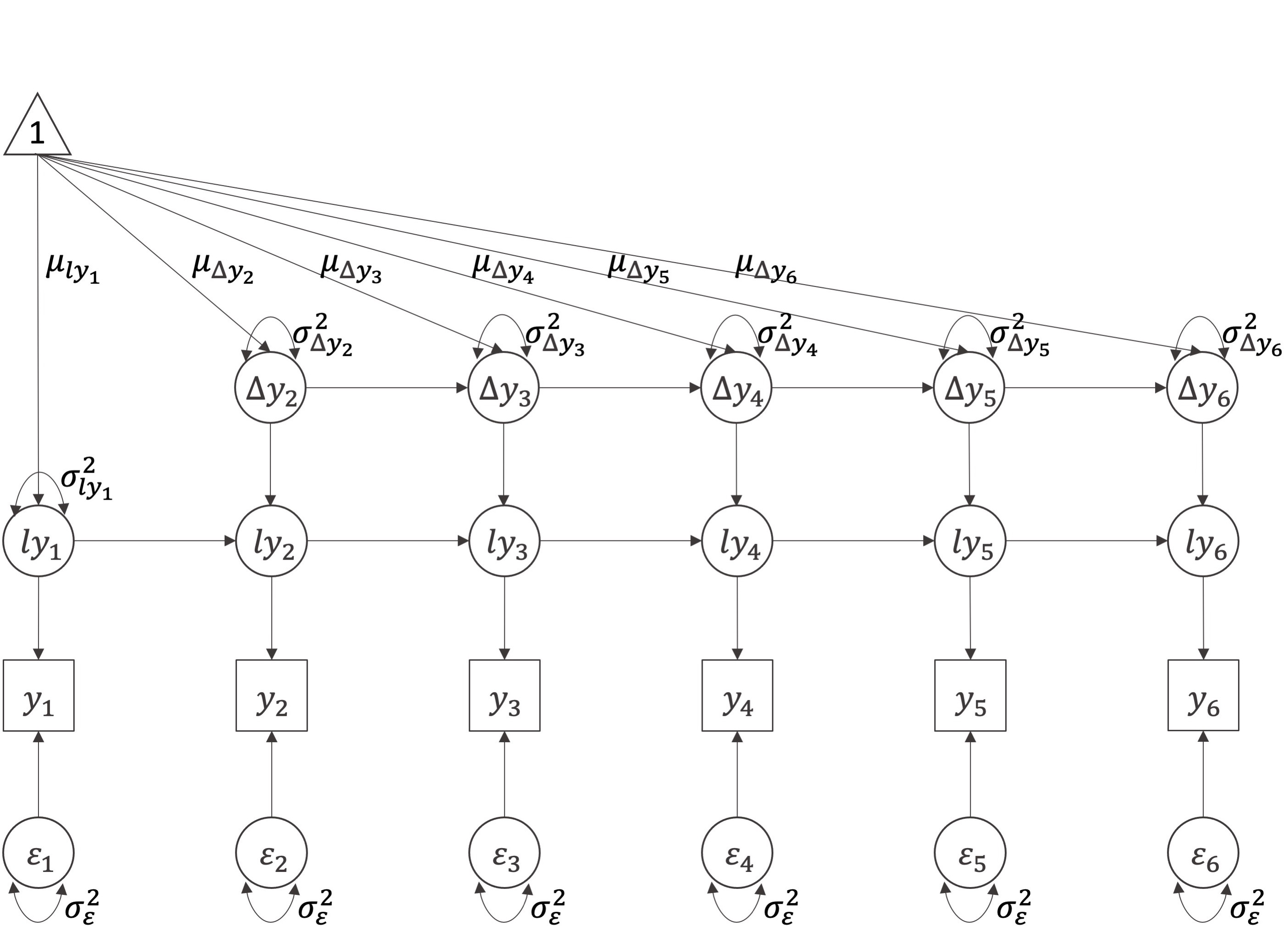}
  \caption{Basic Latent Change Score Model}
  \label{fig:path_LCM}
\end{subfigure}%
\begin{subfigure}{.5\textwidth}
  \centering
  \includegraphics[width=0.95\linewidth]{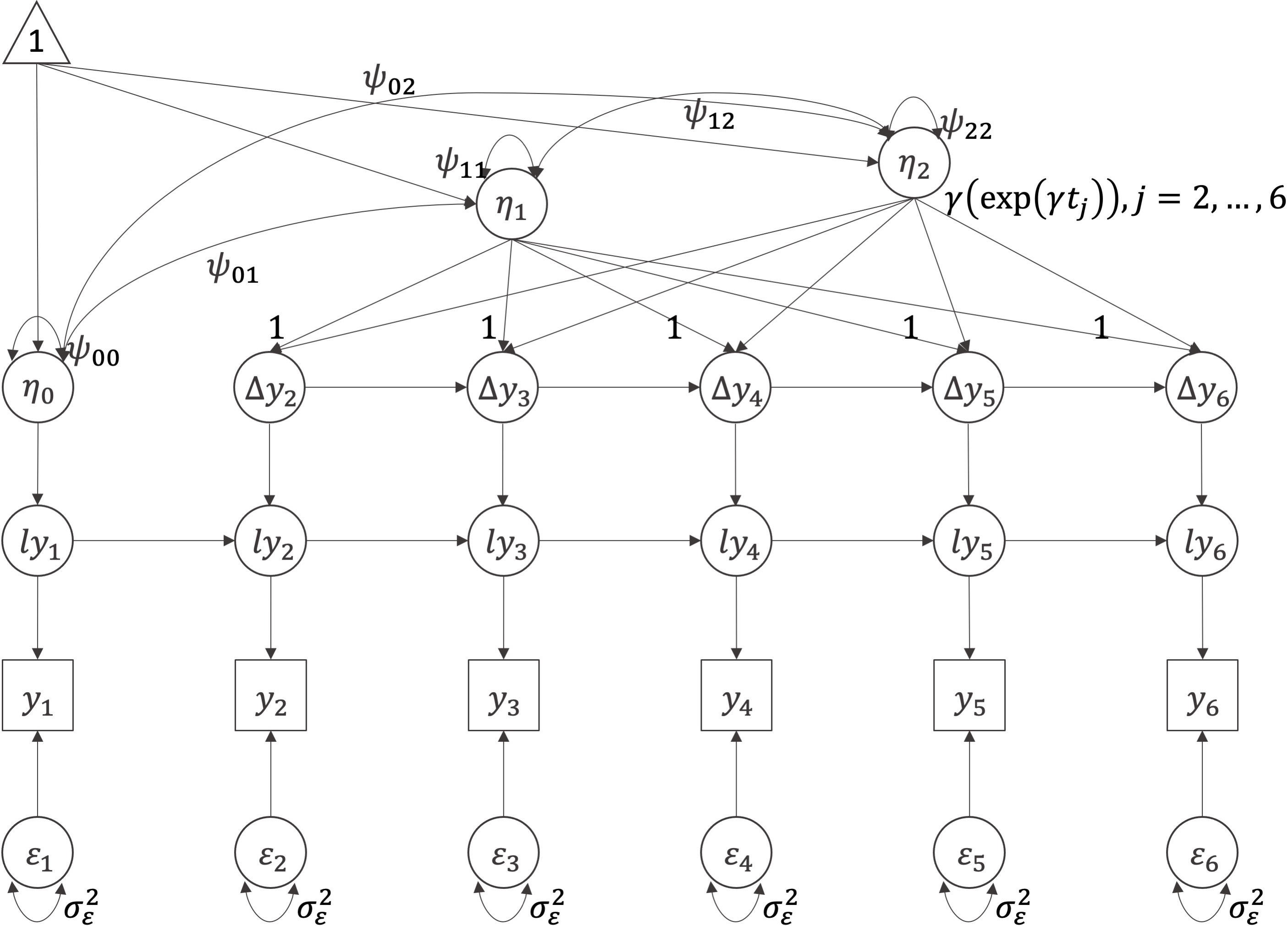}
  \caption{Jenss–Bayley Latent Change Score Model}
  \label{fig:path_JB}
\end{subfigure}
\caption{Path Diagram of the Latent Change Score Models}
\label{fig:LCM_path}
\end{figure}

LCSMs, also referred to as latent difference score models \citep{McArdle2001LCM1, McArdle2001LCM2, McArdle2009LCM}, were developed to integrate difference equations into the structural equation modeling (SEM) framework. In the LCSM, the sequential temporal states of a longitudinal outcome are determined by difference scores. So the LCSM emphasizes the time-dependent change, which is different from the LGCM that represents the time-dependent status. The specification of the LCSM starts from the idea of classical test theory: an individual's score at a specific time point can be viewed as a linear combination of the latent true score and a residual
\begin{equation}\nonumber
y_{ij}=ly_{ij}+\epsilon_{ij},
\end{equation}
where $y_{ij}$, $ly_{ij}$ and $\epsilon_{ij}$ are the observed score, the latent true score, and the residual of the $i^{th}$ individual at time $j$, respectively. The true score at time $j$ is a linear combination of the true score at the prior time point $j-1$ and the latent change score from time $j-1$ to time $j$, which can be written as 
\begin{equation}\nonumber
ly_{ij}=ly_{i(j-1)}+\Delta y_{ij},
\end{equation}
where $ly_{i(j-1)}$ is the latent true score of the $i^{th}$ individual at time $j-1$ and $\Delta y_{ij}$ is the latent change scores from time $j-1$ to time $j$ of the $i^{th}$ individual. We provide the path diagram of the basic LCSM with six repeated measurements in Figure \ref{fig:path_LCM}. With this basic setup of the LCSM, we need to estimate the mean ($\mu_{\Delta y_{j}}$, $j=2,...6$) and variance ($\sigma^{2}_{\Delta y_{j}}$, $j=2,...6$) of each latent change score in addition to the mean ($\mu_{ly_{1}}$) and variance ($\sigma^{2}_{ly_{1}}$) of initial status as well as the residual variance ($\sigma^{2}_{\epsilon}$). We can examine the within-individual changes and between-individual differences in the change scores with their estimated means and variances.

Earlier studies have also shown how to specify and apply a LCSM with a parametric nonlinear growth trajectory. For example, \citet{Grimm2013LCM1} has demonstrated how to obtain a parametric LCSM by taking the first derivative from the corresponding LGCM. \citet[Chapter~18]{Grimm2016growth} has shown that this approach is useful by illustrating the Jenss-Bayley LCSM with an assumption that the ratio of the growth acceleration is roughly similar across all individuals (i.e., only considering the fixed effect of $\gamma$, see Figure \ref{fig:path_JB} for its path diagram with six measurements). From Figure \ref{fig:path_JB}, the intercept (i.e., $\eta_{0}$) is only indicated by the true score at the beginning of the study (we will explain implications of this definition in the LCSM framework in the Discussion section). In addition, $\eta_{1}$ and $\eta_{2}$ together define the latent change scores with factor loadings $1$ and $\gamma\exp(\gamma t_{j})$, respectively. We estimate the mean vector and variance-covariance matrix of the growth factors (i.e., $\eta_{0}$, $\eta_{1}$ and $\eta_{2}$) and the value of the additional parameter $\gamma$ in this specification of the Jenss-Bayley LCSM.

\subsection{Challenges of Implementation of the Jenss-Bayley Latent Change Score Model}\label{I:challenges}
\citet[Chapter~18]{Grimm2016growth} has demonstrated that the Jenss-Bayley LCSM is useful to analyze the longitudinal height data collected as part of the Berkeley Growth Study. The study duration is $36$ months with measures at Month $1$, $3$, $6$, $9$, $12$, $15$, $18$, $24$ and $36$. The existing model successfully addresses the problem of unequally-spaced measurement occasions by specifying a latent change score for each month during the study period. However, researchers still have other challenges when applying this model.

\begin{figure}
\centering
\begin{subfigure}{.5\textwidth}
  \centering
  \includegraphics[width=1\linewidth]{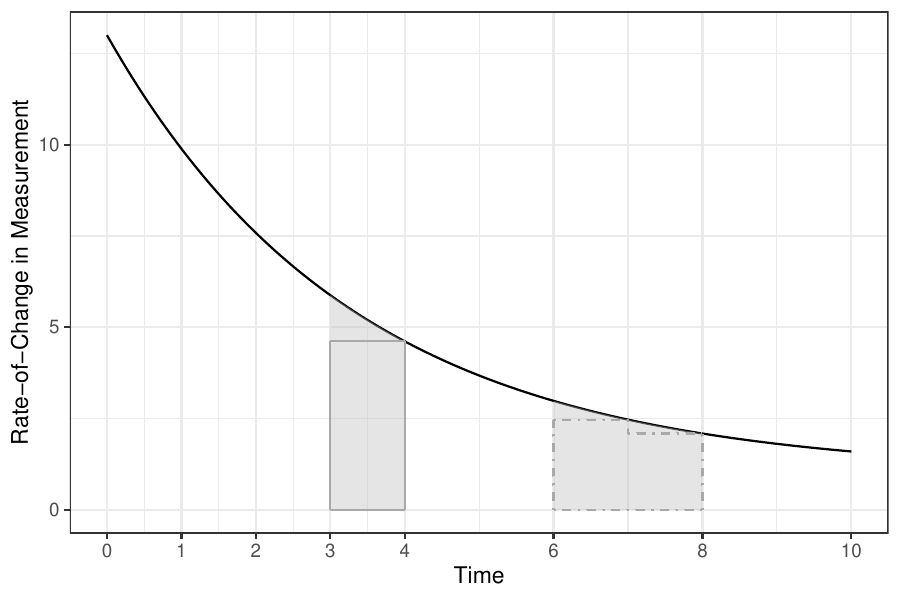}
  \caption{Definition of an Existing Method}
  \label{fig:old}
\end{subfigure}%
\begin{subfigure}{.5\textwidth}
  \centering
  \includegraphics[width=1\linewidth]{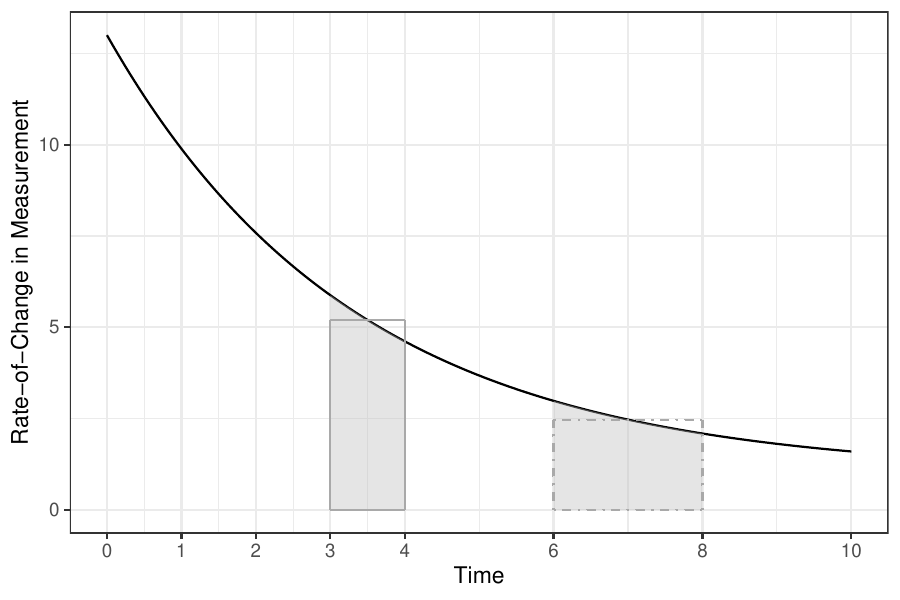}
  \caption{Definition of the Proposed Method}
  \label{fig:new}
\end{subfigure}
\caption{Definition of Latent Change Scores for Jenss-Bayley Latent Change Score Models}
\label{fig:auc}
\end{figure}

First, the latent change score $\Delta y_{ij}$ is the change that occurs in the time interval ($t_{j-1}$, $t_{j}$). Therefore, researchers can only utilize the current model specification, where the instantaneous rate-of-change at $t_{j}$ is used to approximate the change that occurs in the period, with two assumptions (1) the change in the rate-of-change in the interval is infinitesimal and (2) the interval is scaled. To illustrate our point, suppose we have a Jenss-Bayley trajectory, whose the rate-of-change versus time ($r-t$) graph is provided in Figure \ref{fig:old}. The true latent change score from $t=3$ to $t=4$ is the area under the curve (AUC) during that time interval, while the approximated latent change score is the area enclosed by the solid box. From Figure \ref{fig:old}, we can see clearly that the approximation only works well when the change in the instantaneous rate-of-change from $t=3$ to $t=4$ is negligible, and the whole study duration can be rescaled to equal time intervals.

However, this approximation may not be applicable for situations with no evidence supporting an infinitesimal change in the growth rate. In addition, as demonstrated in \citet[Chapter~18]{Grimm2016growth}, where the rescaled time unit is one month, it is feasible to adjust these time intervals to be equal. However, rescaling the intervals to smaller units would complicate the model specification. To address these challenges, we propose to utilize the instantaneous rate-of-change at the mid-point of the time interval ($t_{j-1}$, $t_{j}$) to approximate the average growth rate in that interval. For example, for the latent change score from $t=3$ to $t=4$ in Figure \ref{fig:new}, we use the instantaneous growth rate at $t=3.5$ to approximate the average growth rate; therefore, the approximated latent change score is the area in the solid box in Figure \ref{fig:new}.

Another challenge of longitudinal data analysis is the problem of unstructured measurement occasions, which occurs when time is measured precisely or responses are self-initiated. Earlier studies have demonstrated multiple approaches to address this challenge for different longitudinal models, such as cross-lagged panel model \citep{Voelkle2012ITP}, state-space model \citep{Oud2000ITP}, and growth curve model \citep{Preacher2015repara, Sterba2014individually, Liu2019BLSGM, Liu2021PBLSGM}. In the LCSM framework, \citet{Grimm2018Individually} proposed to specify a latent true score at an individual measurement occasion to obtain the score of each individual. In this article, we propose an alternative method to address the challenge of individual measurement occasions in the LCSM framework by extending the `definition variable' method proposed by \citet{Mehta2000people, Mehta2005people}, where the `definition variable' is an observed variable that adjusts model coefficients to individual-specific values. This method has been widely used in the LGCM framework \citep{Preacher2015repara, Sterba2014individually, Liu2019BLSGM, Liu2021PBLSGM}. In LCSMs, we can define the individual time intervals between two consecutive measurement occasions as the definition variables.

In the novel specification, the latent change score in the time interval ($t_{i(j-1)}$, $t_{ij}$) is defined as the product of the rate-of-change midway through that interval and the length of the interval. In addition to the challenges of time-varying rate-of-change and individual measurement occasions, the novel specification can also solve the issue of unequally-spaced measurement times. Suppose we skip the measurement at $t=7$ as shown in Figure \ref{fig:auc}. With the Jenss-Bayley LCSM in \citet[Chapter~18]{Grimm2016growth}, we have to specify the latent change score for the time interval from $t=6$ to $t=7$ and that from $t=7$ to $t=8$ separately, which are approximated by the instantaneous rate-of-change at $t=7$ and $t=8$, respectively. Using the novel specification, we can approximate the latent change score by the product of the instantaneous rate-of-change at $t=7$ (i.e., the mid-point between $t=6$ and $t=8$) and the time interval from $t=6$ to $t=8$. The approximate value of latent change score from $t=6$ to $t=8$ defined by the existing method and the proposed method is the area enclosed by the dashed box in Figure \ref{fig:old} and \ref{fig:new}, respectively. 

The most important coefficient in the Jenss-Bayley function is the ratio of the growth acceleration $\gamma$, which determines the trajectory shape and the time-taken to the linear asymptote. Given that the time of reaching the level-off stage may be individually different, it may not be reasonable to assume that the ratio of the growth acceleration $\gamma$ is roughly similar across all individuals in practice. In the proposed model, we consider the ratio of the growth acceleration $\gamma$ as the fourth growth factor in addition to $\eta_{0}$, $\eta_{1}$, and $\eta_{2}$ to allow for an individual ratio of the growth acceleration. In the SEM framework, linearization of a target function is one method to obtain an additional growth factor, which can be realized by the Taylor series expansion \citep{Browne1991Taylor, Browne1993Taylor}. Earlier studies, such as \citet{Preacher2012repara}, \citet{Preacher2015repara}, \citet{Liu2019BLSGM}, \citet{Liu2021PBLSGM}, and \citet[Chapter~12]{Grimm2016growth}, have demonstrated how to use this approach for the LGCM. In addition, \citet{Liu2019BLSGM, Liu2021PBLSGM} have shown that the approximation introduced by the Taylor series expansion only affects the model performance slightly by simulation studies. Moreover, \citet{Grimm2013LCM1} has extended the Taylor series expansion to the LCSM framework to estimate an individual ratio of the growth rate. In this article, we use the first-order Taylor series expansion in the LCSM framework and estimate the mean and variance of the ratio of the growth acceleration. 

The proposed model fills existing gaps by demonstrating how to fit a Jenss-Bayley LCSM in the framework of individual measurement occasions to estimate the individual ratio of the growth acceleration and examine the within-individual changes and between-individual differences in the rate-of-change. The remainder of this article is organized as follows. First, in the Method section, we describe the model specification and estimation and demonstrate how to obtain individual rate-of-change for the proposed model. We also introduce a reduced model assuming that the ratio of the growth acceleration is roughly similar across individuals. In the subsequent section, we describe the design of the Monte Carlo simulation to evaluate the proposed model. We then present the performance metrics, including the relative bias, empirical standard error (SE), relative root-mean-squared-error (RMSE), and empirical coverage probability (CP) for a nominal $95\%$ confidence interval of each parameter of interest. In the simulation study, we also examine the statistical power to detect between-individual differences in the ratio of the growth acceleration and compare two approximation methods for the latent change scores. Next, in the Application section, we analyze longitudinal reading achievement scores from the Early Childhood Longitudinal Study, Kindergarten Class $2010-11$ (ECLS-K: $2011$) to demonstrate how to implement the proposed model. We finally discuss practical considerations, methodological considerations, and future directions.

\section{Method}\label{sec:Method}
\subsection{Model Specification}\label{M:specify}
This section describes the Jenss-Bayley LCSM with an unknown random ratio of the growth acceleration in the framework of individual measurement occasions. For the $i^{th}$ individual, we specify the model as
\begin{align}
&y_{ij}=ly_{ij}+\epsilon_{ij},\label{eq:LCSM1}\\
&ly_{ij}=\begin{cases}
\eta_{0i}, & \text{if $j=1$}\\
ly_{i(j-1)}+\Delta y_{ij}, & \text{if $j=2, \dots, J$}
\end{cases},\label{eq:LCSM2}\\
&\Delta y_{ij}\approx dy_{ij}\times(t_{ij}-t_{i(j-1)})\qquad (j=2, \dots, J), \label{eq:LCSM3}\\
&\begin{aligned}
dy_{ij}&=\frac{d}{dt}\big(\eta_{0i}+\eta_{1i}\times t+\eta_{2i}\times(\exp(\gamma_{i}\times t)-1)+\epsilon_{ij}\big)|_{t=t_{ij\_\text{mid}}}\\
&=\eta_{1i}+\eta_{2i}\times\gamma_{i}\times\exp(\gamma_{i}\times t_{ij\_\text{mid}})\qquad (j=2, \dots, J). \label{eq:LCSM4}
\end{aligned}
\end{align}
Equations \ref{eq:LCSM1} and \ref{eq:LCSM2} together define the basic setup for a LCSM, where $y_{ij}$, $ly_{ij}$, and $\epsilon_{ij}$ are the observed measurement, latent true value, and the residual of the $i^{th}$ individual at time $j$, respectively, while $\Delta y_{ij}$ is the true amount of the change that occurs during the time interval ($t_{j-1}$, $t_{j}$) of the individual $i$. For each individual $i$, as shown in Equation \ref{eq:LCSM3}, we further approximate $\Delta y_{ij}$ as the product of the instantaneous rate-of-change midway the time interval (i.e., $dy_{ij}$) and the length of that interval (i.e., $t_{ij}-t_{i(j-1)}$).

We take the first derivative of the Jenss-Bayley trajectory of the $i^{th}$ individual and provide the expression of the instantaneous rate-of-change halfway the time interval ($t_{j-1}$, $t_{j}$) for individual $i$ in Equation \ref{eq:LCSM4}. There are four growth factors in this model specification: $\eta_{0i}$, $\eta_{1i}$, $\eta_{2i}$ and $\exp(\gamma_{i})$ are for the initial status, the slope of the linear asymptote, the vertical distance between two intercepts (i.e., the initial status and the intercept of the linear asymptote), and the ratio of the growth acceleration, respectively. However, the rate-of-change is only determined by $\eta_{1i}$, $\eta_{2i}$ and $\exp(\gamma_{i})$.

\begin{figure}
\centering
  \includegraphics[width=0.88\linewidth]{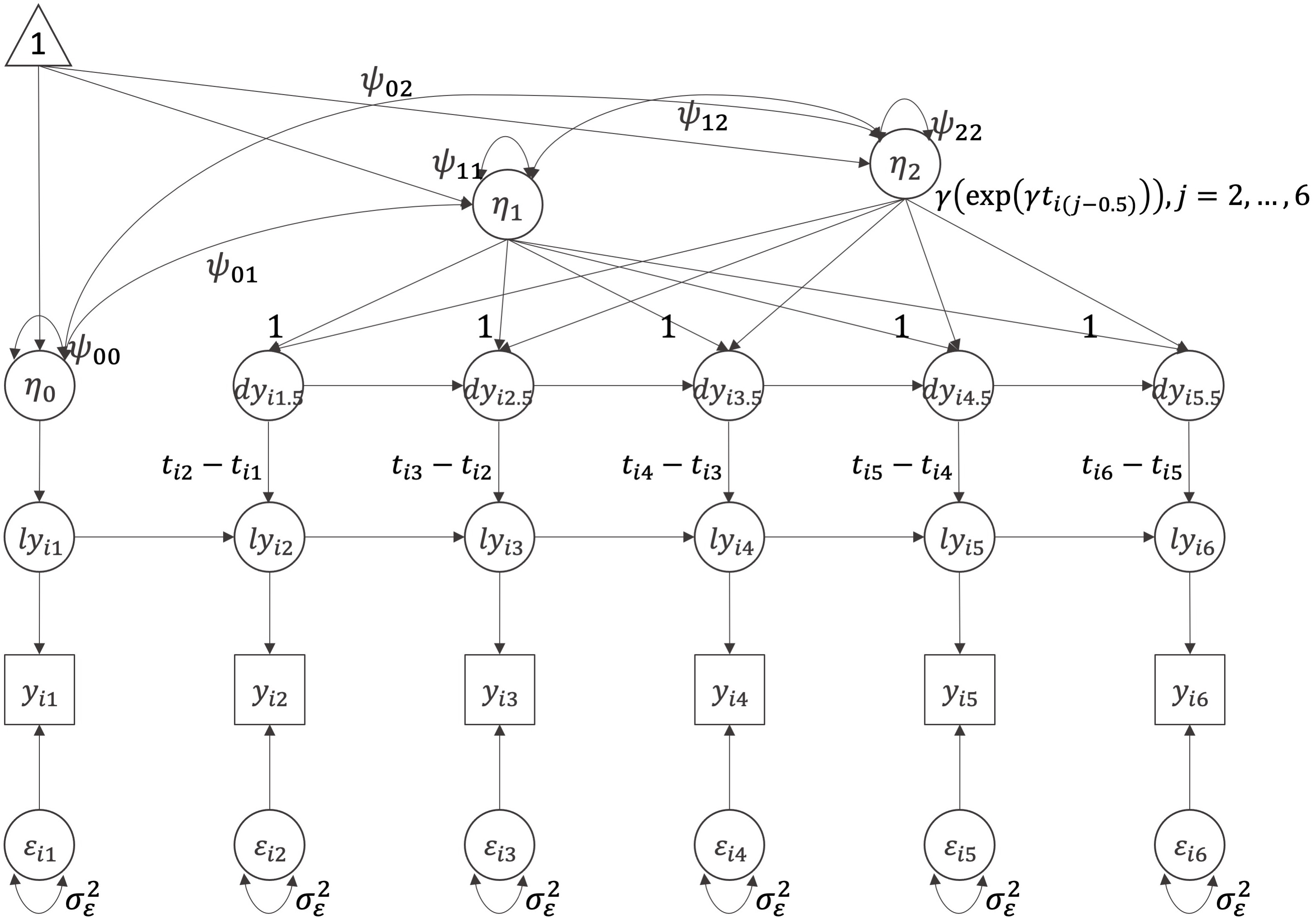}
  \caption{Path Diagram of the Proposed Jenss-Bayley LCSM\\
Note: boxes=manifested variables, circles=latent variables, single arrow=regression paths;
doubled arrow=(co)variances; triangle=constant; diamonds=definition variables.\\
The weights of $\eta_{1}$, $\eta_{2}$, and $\gamma-\mu_{\gamma}$ are $1$, $\mu_{\gamma}\exp{(\mu_{\gamma}t_{j\_\text{mid}})}$ ($j=2,\dots,6$), and $\mu_{\eta_{2}}(1+\mu_{\gamma}t_{j\_\text{mid}})\exp(\mu_{\gamma}t_{j\_\text{mid}})$ ($j=2,\dots,6$), respectively.}
\label{fig:path_novel}
\end{figure}

Note that Equation \ref{eq:LCSM4} does not fit into the LCSM directly since it specifies a nonlinear relationship between the target function $dy_{ij}$ and the growth factor $\gamma_{i}$, which cannot be estimated in the SEM framework directly. \citet{Grimm2013LCM1} has demonstrated how to employ the Taylor series expansion to address the nonlinear relationship in the exponential LCSM. In this article, we employ this method to address this nonlinear relationship in the Jenss-Bayley LCSM. We express $dy_{ij}$ as a linear combination of the growth factors that are related to the rate-of-change (see Appendix \ref{Supp:1a} for details of Taylor series expansion). We provide the path diagram for the novel specification of the Jenss-Bayley LCSM (six measurements) with an unknown random ratio of the growth acceleration in the framework of individual measurement occasions in Figure \ref{fig:path_novel}. Compared to the Jenss-Bayley LCSM specified in Figure \ref{fig:path_JB}, the proposed Jenss-Bayley LCSM has the following characteristics. First, we define the latent instantaneous rate-of-change midway the interval from $t=j-1$ to $t=j$ (i.e., $dy_{ij}$). The weight of the instantaneous growth rate is the length of the individual time interval (i.e., $t_{ij}-t_{i(j-1)}$). Second, in addition to the growth factors $\eta_{0i}$, $\eta_{1i}$ and $\eta_{2i}$, we have an additional growth factor $\gamma_{i}-\mu_{\gamma}$, which is the deviation of an individual $\gamma_{i}$ from the mean value of the logarithmic ratio of the growth acceleration $\mu_{\gamma}$ of individual $i$. From Figure \ref{fig:path_novel}, it is clear that only the growth factors $\eta_{1i}$, $\eta_{2i}$ and $\gamma_{i}-\mu_{\gamma}$ are related to the rate-of-change over time.
Similar to a LGCM, the LCSM specified in Equations \ref{eq:LCSM1}-\ref{eq:LCSM4} has a matrix expression
\begin{equation}\label{eq:LCSM_JB1}
\boldsymbol{y}_{i}\approx\boldsymbol{\Lambda}_{i}\times\boldsymbol{\eta}_{i}+\boldsymbol{\epsilon}_{i},
\end{equation}
where $\boldsymbol{y}_{i}$ is a $J\times1$ vector of the longitudinal outcome of individual $i$ (in which $J$ is the number of measures), $\boldsymbol{\eta}_{i}$ is a $4\times1$ vector of growth factors and can be expressed as
\begin{equation}\nonumber
\boldsymbol{\eta}_{i}=\left(\begin{array}{rrrr}
\eta_{0i} & \eta_{1i} & \eta_{2i} & \gamma_{i}-\mu_{\gamma}
\end{array}\right)^{T},
\end{equation}
of which the first element is the initial status and the other three elements together define the rate-of-change over time. In addition, $\boldsymbol{\Lambda}_{i}$ is a $J\times4$ matrix of the corresponding factor loadings
\noindent
\begin{equation}\nonumber
\tiny
\boldsymbol{\Lambda}_{i}=\begin{pmatrix}
1 & 0 & 0 & 0\\
1 & (t_{i2}-t_{i1}) & \mu_{\gamma}e^{\mu_{\gamma} t_{i2\_mid}}(t_{i2}-t_{i1}) & \mu_{\eta_{2}}e^{\mu_{\gamma} t_{i2\_mid}}(1+\mu_{\gamma}t_{ij\_\text{mid}})(t_{i2}-t_{i1}) \\
1 & \sum_{j=2}^{3}(t_{ij}-t_{i(j-1)}) & \mu_{\gamma}\sum_{j=2}^{3}e^{\mu_{\gamma} t_{ij\_mid}}(t_{ij}-t_{i(j-1)}) & \mu_{\eta_{2}}\sum_{j=2}^{3}e^{\mu_{\gamma}t_{ij\_mid}}(1+\mu_{\gamma}t_{ij\_\text{mid}})(t_{ij}-t_{i(j-1)})\\
\dots & \dots & \dots & \dots \\
1 & \sum_{j=2}^{J}(t_{ij}-t_{i(j-1)}) & \mu_{\gamma}\sum_{j=2}^{J}e^{\mu_{\gamma} t_{ij\_mid}}(t_{ij}-t_{i(j-1)}) & \mu_{\eta_{2}}\sum_{j=2}^{J}e^{\mu_{\gamma}t_{ij\_mid}}(1+\mu_{\gamma}t_{ij\_\text{mid}})(t_{ij}-t_{i(j-1)})\\
\end{pmatrix}.
\end{equation}
The elements of the first column of $\boldsymbol{\Lambda}_{i}$ are the factor loadings of the intercept, the second, third, and fourth columns are the cumulative value of the corresponding slope over time. For example, the second column is the cumulative value of the linear asymptote slope (i.e., $1$) over time, while the third column is the cumulative value of the exponential slope (i.e., $\mu_{\gamma}e^{\mu_{\gamma} t_{ij\_mid}}$). In addition, the elements in the fourth column represent the cumulative value of the `additional slope' that related to the fourth growth factor $\gamma_{i}-\mu_{\gamma}$ (i.e., $\mu_{\eta_{2}}e^{\mu_{\gamma}t_{ij\_mid}}(1+\mu_{\gamma}t_{ij\_\text{mid}})$). Therefore, the product of the grey part of $\boldsymbol{\Lambda}_{i}$ and the corresponding growth factors is interpreted as the amount of the change-from-baseline over time. In addition, the subscript $i$ in $\boldsymbol{\Lambda}_{i}$ suggests the model is defined in the framework of individual measurement occasions. The detailed deviation of obtaining the factor loadings $\boldsymbol{\Lambda}_{i}$ is provided in Appendix \ref{Supp:1b}. Additionally, $\boldsymbol{\epsilon}_{i}$ is a $J\times1$ vector of residuals of individual $i$. The growth factors $\boldsymbol{\eta}_{i}$ can be further expressed as
\begin{equation}\label{eq:LCSM_JB2}
\boldsymbol{\eta}_{i}=\boldsymbol{\mu}_{\boldsymbol{\eta}}+\boldsymbol{\zeta}_{i},
\end{equation}
where $\boldsymbol{\mu}_{\boldsymbol{\eta}}$ is the mean vector of the growth factors, and $\boldsymbol{\zeta}_{i}$ is the vector of deviations from the corresponding mean values of the growth factors.

\subsection{Model Estimation}\label{M:estimate}
To simplify estimation, we assume that the growth factors are normally distributed; that is, $\boldsymbol{\zeta}_{i}\sim \text{MVN}(\boldsymbol{0}, \boldsymbol{\Psi}_{\boldsymbol{\eta}})$, where $\boldsymbol{\Psi}_{\boldsymbol{\eta}}$ is a $4\times4$ variance-covariance matrix of the growth factors. We also assume that residuals are independently and identically normally distributed, that is, for individual $i$, $\boldsymbol{\epsilon}_{i}\sim\text{MVN}(\boldsymbol{0}, \theta_{\epsilon}\boldsymbol{I})$, where $\boldsymbol{I}$ is a $J\times J$ identity matrix. For individual $i$, the expected mean vector and the variance-covariance structure of the repeated outcome of the model given in Equations \ref{eq:LCSM_JB1} and \ref{eq:LCSM_JB2} can be expressed as
\begin{equation}\nonumber
\boldsymbol{\mu}_{i}=\boldsymbol{\Lambda}_{i}\boldsymbol{\mu}_{\boldsymbol{\eta}}
\end{equation}
and
\begin{equation}\nonumber
\boldsymbol{\Sigma}_{i}=\boldsymbol{\Lambda}_{i}\boldsymbol{\Psi}_{\boldsymbol{\eta}}\boldsymbol{\Lambda}_{i}^{T}+\theta_{\epsilon}\boldsymbol{I}.
\end{equation}

The parameters in the model given in Equations \ref{eq:LCSM_JB1} and \ref{eq:LCSM_JB2} include the mean vector and variance-covariance matrix of the growth factors and the residual variance. Therefore, we define
\begin{equation}\nonumber
\boldsymbol{\Theta}_{1}=\{\boldsymbol{\mu}_{\boldsymbol{\eta}}, \boldsymbol{\Psi}_{\boldsymbol{\eta}}, \theta_{\epsilon}\}=\{\mu_{\eta_{0}}, \mu_{\eta_{1}}, \mu_{\eta_{2}}, \mu_{\gamma}, \psi_{00}, \psi_{01}, \psi_{02}, \psi_{0\gamma}, \psi_{11}, \psi_{12}, \psi_{1\gamma}, \psi_{22}, \psi_{2\gamma}, \psi_{\gamma\gamma}, \theta_{\epsilon}\}
\end{equation}
to list the parameters. 

The proposed model is estimated using the full information maximum likelihood (FIML) technique for accounting for the heterogeneity of individual contributions to the likelihood function. The log-likelihood function of each individual and that of the overall sample can be expressed as
\begin{equation}\nonumber
\log lik_{i}(\boldsymbol{\Theta}_{1}|\boldsymbol{y}_{i})=C-\frac{1}{2}\ln|\boldsymbol{\Sigma}_{i}|-\frac{1}{2}\big(\boldsymbol{y}_{i}-\boldsymbol{\mu}_{i})^{T}\boldsymbol{\Sigma}_{i}^{-1}(\boldsymbol{y}_{i}-\boldsymbol{\mu}_{i}\big),
\end{equation}
and
\begin{equation}\nonumber
\log lik(\boldsymbol{\Theta}_{1})=\sum_{i=1}^{n}\log lik_{i}(\boldsymbol{\Theta}_{1}|\boldsymbol{y}_{i}),
\end{equation}
respectively, in which $C$ is a constant, $n$ is the number of individuals, $\boldsymbol{\mu}_{i}$ and $\boldsymbol{\Sigma}_{i}$ are the mean vector and the variance-covariance matrix of the longitudinal outcome $\boldsymbol{y}_{i}$. We construct the proposed model using the R package \textit{OpenMx} with a default optimizer CSOLNP \citep{OpenMx2016package, Pritikin2015OpenMx, Hunter2018OpenMx, User2020OpenMx}. To demonstrate how to fit the proposed Jenss-Bayley LCSM, we provide \textit{OpenMx} code in the online appendix  (\url{https://github.com/Veronica0206/LCSM_projects}). We can also fit the proposed Jenss-Bayley LCSM using other SEM software such as \textit{Mplus} 8. The code is provided on the Github website for researchers who are interested in using it.

\subsection{Obtaining Growth Rate over Time}\label{M:factor_scores}
One objective of employing LCSMs is to obtain the rate-of-change over time, including the mean values and individual scores. This section describes how to calculate these values of the LCSM specified in the framework of individual measurement occasions. The mean values of the growth rate can be calculated from the mean values of the growth factors that define the rate-of-change and the corresponding weight (see Equation \ref{eq:LCSM4_matrix}). Note that the mean values of the growth rate are individual-specific in the framework of individual measurement occasions, since all these individual-specific values are on the same \textit{r-t} curve but correspond to individually-varying time points.

There are multiple approaches to estimate individual scores of the rate-of-change, such as the regression method \citep{Thomson1939score} and the Bartlett method \citep{Bartlett1937score}. In this article, we use the regression method to estimate individual scores. For the $i^{th}$ individual, the joint distribution of repeated measurements $\boldsymbol{y}_{i}$ and all latent variables $\boldsymbol{\eta}_{ai}$ (i.e., $\boldsymbol{\eta}_{ai}=(\begin{array}{ccc}\boldsymbol{\eta}^{T}_{i} & \boldsymbol{ly}^{T}_{i} & \boldsymbol{dy}^{T}_{i}\end{array})^{T}$) is
\begin{equation}\nonumber
\begin{pmatrix}
\boldsymbol{y}_{i} \\ \boldsymbol{\eta}_{ai} 
\end{pmatrix}\sim\text{MVN}\bigg(\begin{pmatrix}
\boldsymbol{\mu}_{i} \\ \boldsymbol{\mu}_{\boldsymbol{\eta}i}
\end{pmatrix}, \begin{pmatrix}
\boldsymbol{\Lambda}_{ai}\boldsymbol{\Psi}_{\boldsymbol{\eta}i}\boldsymbol{\Lambda}^{T}_{ai} + \theta_{\epsilon}\boldsymbol{I} & \boldsymbol{\Lambda}_{ai}\boldsymbol{\Psi}_{\boldsymbol{\eta}i} \\
\boldsymbol{\Psi}_{\boldsymbol{\eta}i}\boldsymbol{\Lambda}^{T}_{ai} & \boldsymbol{\Psi}_{\boldsymbol{\eta}i} 
\end{pmatrix}
\bigg),
\end{equation}
where $\boldsymbol{\mu}_{\boldsymbol{\eta}i}$ and $\boldsymbol{\Psi}_{\boldsymbol{\eta}i}$ are the mean vector and variance-covariance matrix of all latent variables, $\boldsymbol{\Lambda}_{ai}$ is the matrix of the factor loadings from all latent variables to the repeated outcome $\boldsymbol{y}_{i}$ (we provide the deviation and expression of these matrices in Appendix \ref{Supp:1c}).

From the joint distribution, we can calculate the conditional expectation of the individual scores of all the latent variables $\boldsymbol{\eta}_{i}$ given $\boldsymbol{y}_{i}$
\begin{equation}\nonumber
E(\boldsymbol{\eta}_{i}|\boldsymbol{y}_{i})=\boldsymbol{\mu}_{\boldsymbol{\eta}i}+ \boldsymbol{\Psi}_{\boldsymbol{\eta}i}\boldsymbol{\Lambda}^{T}_{ai}(\boldsymbol{\Lambda}_{ai}\boldsymbol{\Psi}_{\boldsymbol{\eta}i}\boldsymbol{\Lambda}^{T}_{ai}+\theta_{\epsilon}\boldsymbol{I})^{-1}(\boldsymbol{y}_{i}-\boldsymbol{\mu}_{i}),
\end{equation}
which suggests the below estimator in which we replace $\boldsymbol{\mu}_{\boldsymbol{\eta}i}$, $\boldsymbol{\Psi}_{\boldsymbol{\eta}i}$, $\boldsymbol{\Lambda}^{T}_{ai}$ and $\boldsymbol{\theta_{\epsilon}}$ with the corresponding estimate
\begin{equation}\nonumber
\hat{\boldsymbol{\eta}}_{i}=\hat{\boldsymbol{\mu}}_{\boldsymbol{\eta}i}+ \hat{\boldsymbol{\Psi}}_{\boldsymbol{\eta}}\hat{\boldsymbol{\Lambda}}^{T}_{ai}(\hat{\boldsymbol{\Lambda}}_{ai}\hat{\boldsymbol{\Psi}}_{\boldsymbol{\eta}i}\hat{\boldsymbol{\Lambda}}^{T}_{ai}+\hat{\theta}_{\epsilon}\boldsymbol{I})^{-1}(\boldsymbol{y}_{i}-\boldsymbol{\mu}_{i}).
\end{equation}
We can estimate $\hat{\boldsymbol{\eta}}_{i}$ by the \textit{OpenMx} function \textit{mxFactorScores()} \citep{OpenMx2016package, Pritikin2015OpenMx, Hunter2018OpenMx, User2020OpenMx, Estabrook2013score}. 

\subsection{Reduced Model}\label{M:Reduced}
We assume that the ratio of the growth acceleration is roughly similar across all individuals and fix the between-individual differences in $\gamma$ to $0$ to build a reduced model
\begin{align}
&\boldsymbol{y}_{i}=\boldsymbol{\Lambda}_{i}\times\boldsymbol{\eta}_{i}+\boldsymbol{\epsilon}_{i}\nonumber\\
&\boldsymbol{\eta}_{i}=\boldsymbol{\mu}_{\boldsymbol{\eta}}+\boldsymbol{\zeta}_{i}\nonumber
\end{align}
where $\boldsymbol{\eta}_{i}$ is a $3\times1$ vector of three growth factors (i.e., the intercept $\eta_{0i}$, the linear asymptote slope $\eta_{1i}$, and the vertical distance between two intercepts $\eta_{2i}$). The corresponding factor loadings $\boldsymbol{\Lambda}_{i}$ also reduces to a $J\times3$ matrix, which can be further expressed as
\begin{equation}\nonumber
\boldsymbol{\Lambda}_{i}=\begin{pmatrix}
1 & 0 & 0\\
1 & (t_{i2}-t_{i1}) & \gamma\times\exp(\gamma\times t_{i2\_mid})\times(t_{i2}-t_{i1})\\
1 & \sum_{j=2}^{3}(t_{ij}-t_{i(j-1)}) & \gamma\times\sum_{j=2}^{3}\exp(\gamma\times t_{ij\_mid})\times(t_{ij}-t_{i(j-1)})\\
\dots & \dots & \dots \\
1 & \sum_{j=2}^{J}(t_{ij}-t_{i(j-1)}) & \gamma\times\sum_{j=2}^{J}\exp(\gamma\times t_{ij\_mid})\times(t_{ij}-t_{i(j-1)})\\
\end{pmatrix}.
\end{equation}
The mean vector and variance-covariance matrix of the growth factors also reduce to a $3\times1$ vector and a $3\times3$ matrix, respectively. We need to estimate the mean vector and variance-covariance matrix of the growth factors, the ratio of growth acceleration, and the residual variance for this reduced model. So
\begin{equation}\nonumber
\boldsymbol{\Theta}_{2}=\{\boldsymbol{\mu}_{\boldsymbol{\eta}}, \gamma, \boldsymbol{\Psi}_{\boldsymbol{\eta}}, \theta_{\epsilon}\}=\{\mu_{\eta_{0}}, \mu_{\eta_{1}}, \mu_{\eta_{2}}, \gamma, \psi_{00}, \psi_{01}, \psi_{02}, \psi_{11}, \psi_{12}, \psi_{22}, \theta_{\epsilon}\}
\end{equation}
lists the parameters. We use the R package \textit{OpenMx} with the optimizer CSOLNP to construct the reduced model and employ the FIML technique to estimate the parameters. We provide the \textit{OpenMx} and \textit{Mplus} 8 syntax on the Github website. 

\section{Model Evaluation}\label{M:Evaluate}
We use a Monte Carlo simulation study to evaluate the proposed model with three goals. The first goal is to evaluate how the approximate value of the latent change score during a time interval and the approximation introduced by the Taylor series expansion affect performance measures, including the relative bias, empirical SE, relative RMSE, and empirical CP of a nominal $95\%$ confidence interval for each parameter. The definitions and estimates of these four performance metrics are given in Table \ref{tbl:metric}. The second goal is to compare the proposed specification of the latent change score with the existing definition; therefore, we generate Jenss-Bayley LGCM-implied data structures and compare the two specifications of the Jenss-Bayley LCSM. Third, we examine the determinants of statistical power to detect between-individual differences in the ratio of the growth acceleration, which is realized by the likelihood-ratio test (LRT). Specifically, we compare the goodness of fit of the proposed Jenss-Bayley LCSM and its reduced version as the two models are nested. The degree of freedom of the LRT is $4$: one is for the variance of the logarithmic ratio of the growth acceleration, and the other three are for its covariances with other growth factors. Additionally, we are interested in examining whether the reduced model performs sufficiently well compared to the full model. 

\begin{table}
\centering
\begin{threeparttable}
\caption{Definitions and Estimates of the Four Performance Metrics}
\begin{tabular}{p{4cm}p{4.5cm}p{5.8cm}}
\hline
\hline
\textbf{Criteria} & \textbf{Definition} & \textbf{Estimate} \\
\hline
Relative Bias & $E_{\hat{\theta}}(\hat{\theta}-\theta)/\theta$ & $\sum_{s=1}^{S}(\hat{\theta}_{s}-\theta)/S\theta$ \\
Empirical SE & $\sqrt{Var(\hat{\theta})}$ & $\sqrt{\sum_{s=1}^{S}(\hat{\theta}_{s}-\bar{\theta})^{2}/(S-1)}$ \\
Relative RMSE & $\sqrt{E_{\hat{\theta}}(\hat{\theta}-\theta)^{2}}/\theta$ & $\sqrt{\sum_{s=1}^{S}(\hat{\theta}_{s}-\theta)^{2}/S}/\theta$ \\
Coverage Probability & $Pr(\hat{\theta}_{\text{lower}}\le\theta\le\hat{\theta}_{\text{upper}})$ & $\sum_{s=1}^{S}I(\hat{\theta}_{\text{lower},s}\le\theta\le\hat{\theta}_{\text{upper},s})/S$\\
\hline
\hline
\end{tabular}
\label{tbl:metric}
\begin{tablenotes}
\small
\item[a] {$\theta$: the population value of the parameter of interest}\\
\item[b] {$\hat{\theta}$: the estimate of $\theta$}\\
\item[c] {$S$: the number of replications and considered as $1,000$ in the simulation study}\\
\item[d] {$s=1,\dots,S$: indexes the simulation replication}\\
\item[e] {$\hat{\theta}_{s}$: the estimate of $\theta$ from the $s^{th}$ replication}\\ 
\item[f] {$\bar{\theta}$: the mean value of $\hat{\theta}_{s}$'s across all $S$ replications}\\
\item[g] {$I()$: an indicator function}
\end{tablenotes}
\end{threeparttable}
\end{table}

Following \citet{Morris2019simulation}, we decided the number of replications $S=1,000$ by an empirical approach in the simulation design. We conducted a pilot simulation run and found that the empirical SEs of all coefficients were less than $0.15$, except for the mean and variance of the initial status and vertical distance between two intercepts (i.e., $\eta_{0}$, $\eta_{2}$, $\psi_{00}$ and $\psi_{22}$). We needed at least $900$ replications to keep the Monte Carlo standard error of the bias\footnote{Bias is the most important performance measure, and the formula for its Monte Carlo standard error is $\text{Monte Carlo SE(Bias)}=\sqrt{Var(\hat{\theta})/S}$ \citep{Morris2019simulation}.} below $0.005$. Accordingly, out of more conservative consideration, we decided to proceed with $S=1,000$.

\subsection{Design of Simulation Study}\label{Simu:design}
As mentioned earlier, the parameters of the most interest in the full model are the mean ($\mu_{\gamma}$) and variance ($\psi_{\gamma\gamma}$) of the logarithmic ratio of the growth acceleration. The conditions hypothesized to influence the estimation of these $\gamma$-related parameters and other model parameters are the number of repeated measurements, whether the measurement occasions are equally spaced, the trajectory shape, the variance of the ratio of the growth acceleration, the measurement precision, and the sample size. Accordingly, we did not examine the conditions that presumably would not affect the model performance meaningfully. For example, the distribution of the initial status, which only affects the position of a trajectory, does not impact the rate-of-change. So we fixed its distribution and kept its index of dispersion ($\sigma^{2}/\mu$) at $0.3$ (i.e., a one-tenth scale). As shown in Figure \ref{fig:JB_LGC}, a reasonable range of $\gamma_{i}$ in a Jenss-Bayley growth curve is ($-0.9$, $-0.5$); accordingly, we took their average $-0.7$ as the mean value of the logarithmic ratio of the growth acceleration. Additionally, we fixed the distribution of the vertical difference between the two intercepts and set the growth factors positively correlated to a moderate level ($\rho=0.3$).

All conditions that we considered in the simulation design are provided in Table \ref{tbl:simu_design}. For the proposed model, one factor of interest is the number of repeated measurements. In general, a model for analyzing longitudinal data should perform better if we have more follow-up times \citep{Timmons2015timing}. We want to examine whether this is the case with the proposed model. Additionally, following \citet{Timmons2015timing}, we want to examine whether the measurement occasions are equally-placed or not would affect the model performance, given that the proposed model's rate-of-change is not constant. To this end, we selected two different levels of the number of measurement occasions: seven and ten, assuming that the study duration is the same across conditions. As shown in Table \ref{tbl:simu_design}, for the conditions with seven measurements, we considered equidistant waves, while for the conditions with ten measurements, we set them to be equally spaced or placed more measurements at the early phase of the study since the initial development of the Jenss-Bayley growth curve is steep, as shown in Figure \ref{fig:JB_LCM}. We then set the time window around each wave at a medium level $(-0.25, +0.25)$ for individual measurement occasions \citep{Coulombe2015ignoring}. 

\begin{table}
\centering
\resizebox{1.0\textwidth}{!}{
\begin{threeparttable}
\setlength{\tabcolsep}{4pt}
\renewcommand{\arraystretch}{0.6}
\caption{Simulation Design for the Jenss-Bayley Latent Change Score Model in the Framework of Individual Measurement Occasions}
\begin{tabular}{p{11cm} p{13.2cm}}
\hline
\hline
\multicolumn{2}{c}{\textbf{Fixed Conditions}}\\
\hline
\textbf{Variables} & \textbf{Conditions} \\
\hline
Distribution of the Initial Status (True Intercept) & $\eta_{0i}\sim N(50, 4^{2})$ (i.e., $\mu_{\eta_{0}}=50$, $\psi_{00}=16$) \\
Distribution of the Vertical Distance between Two Intercepts\tnote{a} & $\eta_{2i}\sim N(-30, 6^{2})$ (i.e., $\mu_{\eta_{2}}=-30$, $\psi_{22}=36$) \\
Mean of the Log-value of Ratio of the Growth Acceleration & $\mu_{\gamma}=-0.7$ (i.e., $\exp(\mu_{\gamma})=0.5$) \\
Correlations of Growth Factors\tnote{b} & $\rho=0.3$ \\
\hline
\hline
\multicolumn{2}{c}{\textbf{Manipulated Conditions (Full Factorial)}}\\
\hline
\textbf{Variables} & \textbf{Conditions} \\
\hline
\multirow{3}{*}{Time ($t_{j}$)} & $7$ equally-spaced: $t_{j}=0, 1.50, 3.00, 4.50, 6.00, 7.50, 9.00$\\
& $10$ equally-spaced: $t_{j}=0, 1.00, 2.00, 3.00, 4.00, 5.00, 6.00, 7.00, 8.00, 9.00$\\
& $10$ unequally-spaced: $t_{j}=0, 0.75, 1.50, 2.25, 3.00, 3.75, 4.50, 6.00, 7.50, 9.00$\\
\hline
Individual $t_{ij}$ & $t_{ij} \sim U(t_{j}-\Delta, t_{j}+\Delta)$ ($\Delta=0.25$) \\
\hline
\multirow{2}{*}{Sample Size} & $n=200$ \\
& $n=500$ \\
\hline
\multirow{2}{*}{Distribution of the Slope of the Linear Asymptote}
& $\eta_{1i}\sim N(2.5, 1.0^{2})$ (i.e., $\mu_{\eta_{1}}=2.5$, $\psi_{11}=1.00$)\\
& $\eta_{1i}\sim N(1.0, 0.4^{2})$ (i.e., $\mu_{\eta_{1}}=1.0$, $\psi_{11}=0.16$)\\
\hline
\multirow{3}{*}{Variance of the Log-value of Ratio of the Growth Acceleration} 
& $\psi_{\gamma\gamma}=0$ \\
& $\psi_{\gamma\gamma}=0.05^{2}$ (i.e., $95\%$ of $\exp(\gamma_{i})$ are in the interval of ($0.45$, $0.55$).)\\
& $\psi_{\gamma\gamma}=0.10^{2}$ (i.e., $95\%$ of $\exp(\gamma_{i})$ are in the interval of ($0.41$, $0.61$).)\\
\hline
\multirow{2}{*}{Residual Variance} & $\theta_{\epsilon}=1$ \\
& $\theta_{\epsilon}=2$ \\
\hline
\hline
\end{tabular}
\label{tbl:simu_design}
\begin{tablenotes}
\small
\item[a] {Two intercepts mean the actual intercept (i.e., the initial status) and the linear asymptote intercept.}\\
\item[b] {In the simulation design, by `Growth Factors', we mean the initial status, the slope of the linear asymptote, the vertical distance between two intercepts, and the log-value of ratio of the growth acceleration.}
\end{tablenotes}
\end{threeparttable}}
\end{table}

Additionally, we set the standard deviation of the logarithmic ratio of the growth acceleration as $0$, $0.05$, and $0.10$ as zero, medium, and large individual differences. We chose these values to keep $95\%$ individual ratios in the range shown in Figure \ref{fig:JB_LGC}. With the three levels of magnitude of between-individual differences, we aim to examine whether and how the approximation introduced by the Taylor series expansion affects model performance. We also investigated other coefficients that affect the trajectory shape, which may then affect the model performance. For example, we considered two different distributions for the slope of the linear asymptote: $N(2.5, 1.0^{2})$ and $N(1.0, 0.4^{2})$ for a large and small growth rate in the later development. We adjusted the distribution variance so that the slope of $95\%$ individuals is positive. In addition,  we considered two levels of measurement precision and two levels of sample size as shown in Table \ref{tbl:simu_design}. 

\subsection{Data Generation and Simulation Step}\label{evaluation:step}
For each condition listed in Table \ref{tbl:simu_design}, we conducted the simulation study for the proposed model according to the general steps outlined as follows:
\begin{enumerate}
\item Generate data for the growth factors using the R package \textit{MASS} \citep{Venables2002Statistics},
\vspace{-2mm}
\item Generate the time structure with $J$ waves $t_{j}$ as specified in Table \ref{tbl:simu_design} and obtain individual measurement occasions: $t_{ij}\sim U(t_{j}-\Delta, t_{j}+\Delta)$ ($\Delta=0.25$),
\vspace{-2mm}
\item Calculate factor loadings, which are functions of individual measurement occasions and the ratio of the growth acceleration, for each individual,
\vspace{-2mm}
\item Generate the Jenss-Bayley LGCM-implied data structures based on growth factors, factor loadings, and residual variances,
\vspace{-2mm}
\item Implement the full model with the novel specification and that with the existing specification and the corresponding reduced model on the generated data, estimate the parameters, and construct corresponding $95\%$ Wald CIs,
\vspace{-2mm}
\item Repeat the above steps until achieving $1,000$ convergent solutions.
\end{enumerate}

\section{Result}\label{sec:Result}
\subsection{Model Convergence and Proper Solution}\label{R:Preliminary}
We first examined the convergence\footnote{The convergence is defined as achieving \textit{OpenMx} status code $0$, indicating a successful optimization, until up to $10$ trials with different sets of initial values \citep{OpenMx2016package}.} rate and the proportion of improper solutions for each condition before evaluating the performance of the proposed Jenss-Bayley LCSM. The proposed model and its reduced version converged satisfactorily: all $72$ conditions reported a $100\%$ convergence rate of the full and reduced Jenss-Bayley LCSM.

Following \citet[Chapter~2]{Bollen2005LGC}, we investigated the pattern of `improper solutions' (referring to the estimates that are impossible in the population), including negative estimated variances of growth factors and/or out-of-range correlations (i.e., beyond $[-1, 1]$) between growth factors. Table \ref{tbl:Improper} presents the occurrence of improper solutions produced by the proposed Jenss-Bayley LCSM under all conditions. The improper solutions include negative variances of the logarithmic ratio of the growth acceleration or its out-of-range correlation with other growth factors. From the table, the proposed model is capable of producing proper solutions when we correctly specify the model (i.e., employ the model under the conditions with a non-zero standard deviation of the logarithmic ratio of the growth acceleration), though the number of improper solutions was relatively large if the model was over-specified under the conditions where the standard deviation of the logarithmic ratio was $0$. Therefore, we replaced the estimates from the full Jenss-Bayley LCSM with the estimated values from the reduced version for model evaluation when such improper solutions occurred.

\begin{table}
\centering
\resizebox{1.0\textwidth}{!}{
\begin{threeparttable}
\setlength{\tabcolsep}{5pt}
\renewcommand{\arraystretch}{0.75}
\caption{Number of Improper Solutions among $1,000$ Replications of the Proposed Model}
\begin{tabular}{p{5.0cm}|p{3.0cm}|p{2.5cm}|R{1.5cm}R{1.5cm}|R{1.5cm}R{1.5cm}}
\hline
\hline
& & & \multicolumn{2}{c}{$\theta_{\epsilon}=1$} & \multicolumn{2}{c}{$\theta_{\epsilon}=2$}\\ 
\hline
& & & \textbf{$n=200$} & \textbf{$n=500$} & \textbf{$n=200$} & \textbf{$n=500$}\\
\hline
\hline
\multirow{6}{*}{Ten Equally-spaced Wave}
& \multirow{3}{*}{$\eta_{1i}\sim N(2.5, 1.0^2)$}
& $sd(\gamma)=0$ & $521//42$\tnote{a} & $513//25$ & $538//46$ & $479//30$ \\
& & $sd(\gamma)=0.05$ & $63//69$ & $9//15$ & $221//102$ & $99//70$ \\
& & $sd(\gamma)=0.10$ & $0//0$ & $0//0$ & $3//23$ & $0//0$ \\
\cline{2-7}
&\multirow{3}{*}{{$\eta_{1i}\sim N(1.0, 0.4^2)$}} 
& $sd(\gamma)=0$ & $528//48$ & $478//26$ & $507//60$ & $495//22$ \\
& & $sd(\gamma)=0.05$ & $57//64$ & $5//27$ & $203//115$ & $92//59$ \\
& & $sd(\gamma)=0.10$ & $0//2$ & $0//0$ & $2//41$ & $0//2$ \\
\hline
\multirow{6}{*}{Ten Unequally-spaced Wave}
& \multirow{3}{*}{$\eta_{1i}\sim N(2.5, 1.0^2)$}
& $sd(\gamma)=0$ & $536//47$ & $517//29$ & $514//45$ & $492//24$ \\
& & $sd(\gamma)=0.05$ & $51//62$ & $5//10$ & $204//84$ & $81//76$ \\
& & $sd(\gamma)=0.10$ & $0//1$ & $0//0$ & $2//18$ & $0//0$ \\
\cline{2-7}
&\multirow{3}{*}{{$\eta_{1i}\sim N(1.0, 0.4^2)$}} 
& $sd(\gamma)=0$ & $523//54$ & $521//31$ & $489//44$ & $522//27$ \\
& & $sd(\gamma)=0.05$ & $42//68$ & $3//13$ & $200//118$ & $63//81$ \\
& & $sd(\gamma)=0.10$ & $0//0$ & $0//0$ & $0//36$ & $0//2$ \\
\hline
\multirow{6}{*}{Seven Equally-spaced Wave}
& \multirow{3}{*}{$\eta_{1i}\sim N(2.5, 1.0^2)$}
& $sd(\gamma)=0$ & $531//46$ & $484//31$ & $517//36$ & $478//27$ \\
& & $sd(\gamma)=0.05$ & $160//92$ & $45//57$ & $309//79$ & $193//79$ \\
& & $sd(\gamma)=0.10$ & $0//14$ & $0//0$ & $25//103$ & $2//16$ \\
\cline{2-7}
&\multirow{3}{*}{{$\eta_{1i}\sim N(1.0, 0.4^2)$}} 
& $sd(\gamma)=0$ & $503//48$ & $471//28$ & $517//49$ & $504//32$ \\
& & $sd(\gamma)=0.05$ & $158//102$ & $40//64$ & $305//89$ & $183//90$ \\
& & $sd(\gamma)=0.10$ & $0//24$ & $0//0$ & $30//103$ & $0//26$ \\
\hline
\hline
\end{tabular}
\label{tbl:Improper}
\begin{tablenotes}
\small
\item[a] {$521//42$ suggests that, for the proposed model, among $1,000$ replications with convergent solutions, we have $521$ and $42$ improper solutions result from negative variances of logarithmic ratio of the growth acceleration and its out-of-range correlations with other growth factors, respectively.}
\end{tablenotes}
\end{threeparttable}}
\end{table}

\subsection{Performance of the Proposed Jenss-Bayley Latent Change Score Model}\label{R:Primary}
This section summarizes the performance measures for the simulation study, including the relative bias, empirical SE, relative RMSE, and empirical CP for a nominal $95\%$ confidence interval for each parameter. Generally, the proposed Jenss-Bayley LCSM can provide unbiased and accurate point estimates with the target coverage probabilities of a nominal $95\%$ confidence interval. We first provide the median and range of each performance measure for each parameter of interest across conditions given the size of parameters and simulation conditions. We then discuss how the simulation conditions affect model performance. 

Tables \ref{tbl:rBias} and \ref{tbl:empSE} present the median and range of the relative bias and empirical SE of each parameter of interest across the conditions with ten repeated measurements for the proposed Jenss-Bayley LCSM and the reduced model. We first obtained the relative bias/empirical SE of the $1,000$ replications under each condition and then summarized them as the corresponding median and range across all conditions.

\begin{table}
\centering
\begin{threeparttable}
\setlength{\tabcolsep}{5pt}
\renewcommand{\arraystretch}{0.75}
\caption{Median and range of the Relative Bias of Parameters in the Jenss-Bayley LCSMs ($10$ Repeated Measurements)}
\begin{tabular}{p{3.0cm}p{1cm}R{5.5cm}R{5.5cm}}
\hline
\hline
\multicolumn{4}{c}{\textbf{Proposed Expression of Latent Change Scores}} \\
\hline
\hline
& \textbf{Para.} & \textbf{Reduced Model} & \textbf{Full Model} \\
\hline
& & Median (Range) & Median (Range) \\
\hline
\multirow{4}{*}{\textbf{\makecell[l]{Growth Factor \\ Means}}} 
& $\mu_{\eta_{0}}$ & $-0.0003$ ($-0.0008$, $0.0003$) & $0.0000$ ($-0.0006$, $0.0007$) \\
& $\mu_{\eta_{1}}$ & $0.0132$ ($-0.0046$, $0.0694$) & $0.0066$ ($-0.0048$, $0.0414$) \\
& $\mu_{\eta_{2}}$ & $0.0101$ ($-0.0046$, $0.0236$) & $0.0131$ ($0.0013$, $0.0238$) \\
& $\mu_{\gamma}$ & $0.0095$ ($-0.0027$, $0.0272$) & $0.0040$ ($-0.0027$, $0.0118$) \\
\hline
\multirow{4}{*}{\textbf{\makecell[l]{Growth Factor \\ Variances}}} 
& $\psi_{00}$ & $-0.0255$ ($-0.0551$, $-0.0004$) & $-0.0022$ ($-0.0093$, $0.0085$) \\
& $\psi_{11}$ & $0.1293$ ($-0.0117$, $0.4365$) & $0.0066$ ($-0.0060$, $0.0466$) \\
& $\psi_{22}$ & $0.1388$ ($0.0154$, $0.3098$) & $0.0431$ ($0.0219$, $0.0732$) \\
& $\psi_{\gamma\gamma}$ & ---\tnote{a} & $0.1569$ ($-0.0356$, NA)\tnote{b} \\
\hline
\hline
\multicolumn{4}{c}{\textbf{Existing Expression of Latent Change Scores}} \\
\hline
\hline
& \textbf{Para.} & \textbf{Reduced Model} & \textbf{Full Model} \\
\hline
& & Median (Range) & Median (Range) \\
\hline
\multirow{4}{*}{\textbf{\makecell[l]{Growth Factor \\ Means}}} 
& $\mu_{\eta_{0}}$ & $-0.0003$ ($-0.0010$, $0.0004$) & $0.0000$ ($-0.0005$, $0.0005$) \\
& $\mu_{\eta_{1}}$ & $0.0232$ ($-0.0300$, $0.1069$) & $0.0147$ ($-0.0301$, $0.0829$) \\
& $\mu_{\eta_{2}}$ & $0.3900$ ($0.3138$, $0.4636$) & $0.3915$ ($0.3163$, $0.4641$) \\
& $\mu_{\gamma}$ & $-0.0034$ ($-0.019$, $0.0183$) & $-0.0071$ ($-0.0187$, $0.0050$) \\
\hline
\multirow{4}{*}{\textbf{\makecell[l]{Growth Factor \\ Variances}}} 
& $\psi_{00}$ & $-0.0260$ ($-0.0547$, $0.0004$) & $-0.0052$ ($-0.0125$, $0.0022$) \\
& $\psi_{11}$ & $0.1271$ ($-0.0172$, $0.4322$) & $0.0088$ ($-0.0108$, $0.0442$) \\
& $\psi_{22}$ & $1.1965$ ($0.7503$, $1.7463$) & $1.1205$ ($0.7447$, $1.6868$) \\
& $\psi_{\gamma\gamma}$ & --- & $0.5260$ ($-0.0321$, NA) \\
\hline
\hline
\end{tabular}
\label{tbl:rBias}
\begin{tablenotes}
\small
\item[a] {--- indicates that the relative biases are not available from the reduced model.}\\
\item[b] {NA indicates that the bound of relative bias is not available. The model performance under the conditions with $0$ population value of the variance of the logarithmic ratio of the growth acceleration is of interest where the relative bias would go infinity. The median (range) of the bias of the logarithmic ratio of the growth acceleration for the proposed expression and existing expression is $0.0004$ ($-0.0004$, $0.0024$) and $0.0011$ ($-0.0003$, $0.0028$), respectively. }
\end{tablenotes}
\end{threeparttable}
\end{table}

\begin{table}
\centering
\begin{threeparttable}
\setlength{\tabcolsep}{5pt}
\renewcommand{\arraystretch}{0.75}
\caption{Median and range of the Empirical SE of Parameters in the Jenss-Bayley LCSMs ($10$ Repeated Measurements)}
\begin{tabular}{p{3.0cm}p{1cm}R{5.5cm}R{5.5cm}}
\hline
\hline
\multicolumn{4}{c}{\textbf{Proposed Expression of Latent Change Scores}} \\
\hline
\hline
& \textbf{Para.} & \textbf{Reduced Model} & \textbf{Full Model} \\
\hline
& & Median (Range) & Median (Range) \\
\hline
\multirow{4}{*}{\textbf{\makecell[l]{Growth Factor \\ Means}}} 
& $\mu_{\eta_{0}}$ & $0.2362$ ($0.1811$, $0.3178$) & $0.2385$ ($0.1764$, $0.3173$) \\
& $\mu_{\eta_{1}}$ & $0.0501$ ($0.0319$, $0.0825$) & $0.0499$ ($0.0320$, $0.0836$) \\
& $\mu_{\eta_{2}}$ & $0.3882$ ($0.2871$, $0.5175$) & $0.3914$ ($0.2750$, $0.5465$) \\
& $\mu_{\gamma}$ & $0.0082$ ($0.0047$, $0.0142$) & $0.0083$ ($0.0048$, $0.0141$) \\
\hline
\multirow{4}{*}{\textbf{\makecell[l]{Growth Factor \\ Variances}}} 
& $\psi_{00}$ & $1.3842$ ($0.9689$, $1.7838$) & $1.4098$ ($1.0028$, $1.8618$) \\
& $\psi_{11}$ & $0.0651$ ($0.0274$, $0.1335$) & $0.0686$ ($0.0318$, $0.1286$) \\
& $\psi_{22}$ & $3.6927$ ($2.4169$, $5.1340$) & $3.7658$ ($2.5485$, $5.4387$) \\
& $\psi_{\gamma\gamma}$ & ---\tnote{a} & $0.0014$ ($0.0000$, $0.0036$) \\
\hline
\hline
\multicolumn{4}{c}{\textbf{Existing Expression of Latent Change Scores}} \\
\hline
\hline
& \textbf{Para.} & \textbf{Reduced Model} & \textbf{Full Model} \\
\hline
& & Median (Range) & Median (Range) \\
\hline
\multirow{4}{*}{\textbf{\makecell[l]{Growth Factor \\ Means}}} 
& $\mu_{\eta_{0}}$ & $0.2362$ ($0.1810$, $0.3174$) & $0.2367$ ($0.1808$, $0.3170$) \\
& $\mu_{\eta_{1}}$ & $0.0497$ ($0.0318$, $0.0830$) & $0.0495$ ($0.0319$, $0.0834$) \\
& $\mu_{\eta_{2}}$ & $0.5264$ ($0.3584$, $0.7592$) & $0.5261$ ($0.3524$, $0.7545$) \\
& $\mu_{\gamma}$ & $0.0083$ ($0.0048$, $0.0140$) & $0.0083$ ($0.0049$, $0.0140$) \\
\hline
\multirow{4}{*}{\textbf{\makecell[l]{Growth Factor \\ Variances}}} 
& $\psi_{00}$ & $1.3889$ ($0.9702$, $1.7867$) & $1.4158$ ($1.0019$, $1.8584$) \\
& $\psi_{11}$ & $0.0657$ ($0.0275$, $0.1343$) & $0.0676$ ($0.0325$, $0.1299$) \\
& $\psi_{22}$ & $7.1120$ ($4.2073$, $10.9451$) & $7.1392$ ($4.2064$, $11.0072$) \\
& $\psi_{\gamma\gamma}$ & --- & $0.0016$ ($0.0000$, $0.0036$) \\
\hline
\hline
\end{tabular}
\label{tbl:empSE}
\begin{tablenotes}
\small
\item[a] {--- indicates that the empirical standard errors are not available from the reduced model.}
\end{tablenotes}
\end{threeparttable}
\end{table}

From Tables \ref{tbl:rBias} and \ref{tbl:empSE}, we can see that the proposed Jenss-Bayley LCSM with the novel specification generally provided unbiased point estimates and small empirical SEs. Specifically, for the proposed Jenss-Bayley LCSM, the magnitude of relative biases of the growth factor means was below $0.05$, and that of the variance of the initial status ($\psi_{00}$), slope asymptote ($\psi_{11}$) and distance between two intercepts ($\psi_{22}$) was less than $0.08$. On the other hand, from Table \ref{tbl:rBias}, the proposed model may produce biased estimates for the variance of the logarithmic ratio of the growth acceleration ($\psi_{\gamma\gamma}$): the median value of relative biases of $\psi_{\gamma\gamma}$ was $0.16$. Additionally, fewer measurements increased the relative biases slightly (the summary of the relative bias of each parameter under the conditions with seven repeated measurements is provided in Table \ref{tbl:Metric7} in Appendix \ref{Supp:2}).

\begin{figure}
\centering
\includegraphics[width=0.91\linewidth]{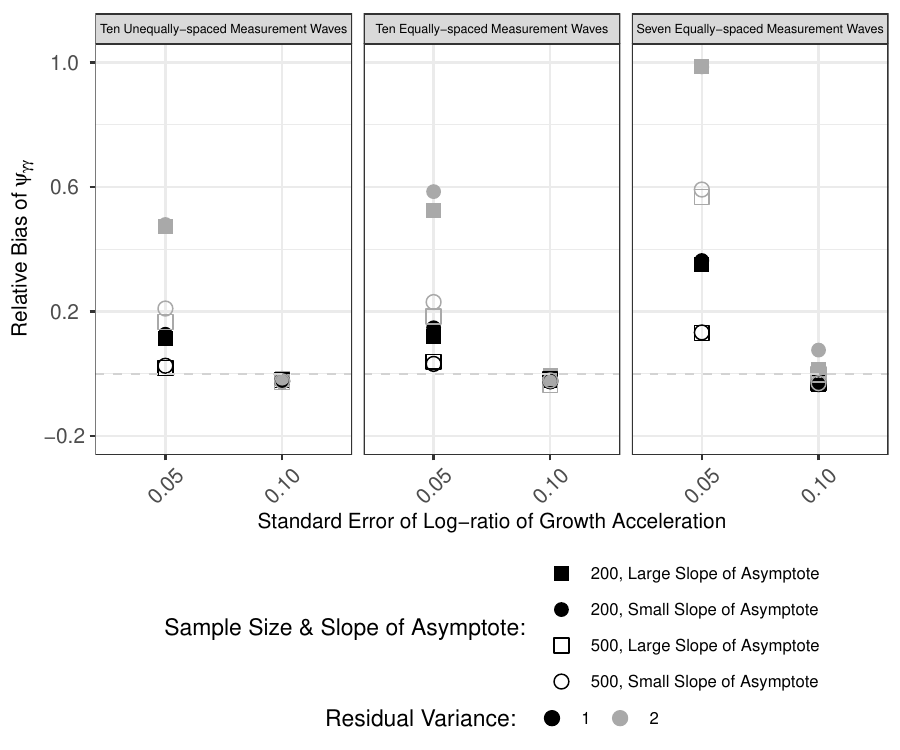}
\caption{Relative Biases of Variances of Logarithmic Ratio of the Growth Acceleration}
\label{fig:rBias}
\end{figure}

We then plot the relative bias under each condition for $\psi_{\gamma\gamma}$ in Figure \ref{fig:rBias}, from which we can observe the influence on these estimates of the conditions we considered in the simulation design. First, relative biases were small under the conditions where the standard deviation of the logarithmic ratio of the growth acceleration was $0.10$. Second, $\psi_{\gamma\gamma}$ was over-estimated when the standard deviation was set to $0.05$. Third, the relative biases were smaller under the conditions with more repeated measurements (i.e., ten in our case); more importantly, taking more measurements in the early stage of the study would further decrease the bias.

As shown in Table \ref{tbl:empSE}, the estimates from the proposed Jenss-Bayley LCSM and its reduced model were precise: the magnitude of empirical SEs of the parameters related to the slope or logarithmic ratio of the growth acceleration was less than $0.15$, although the empirical SEs of other parameters were relatively large: the empirical SE of the mean value of the initial status and that of the vertical distance between two intercepts was around $0.4$, while the empirical SE of the variance of the initial status and that of the vertical distance between two intercepts was around $1.4$ and $3.8$, respectively. Those relatively large empirical SEs were due to the large scale of their population values. 

We provide the median and range of relative RMSE of each parameter for the proposed model and its reduced version under the conditions with ten repeated measures in Table \ref{tbl:rRMSE}. The relative RMSE combines bias and precision to examine the point estimate holistically. From the table, the magnitude of relative RMSEs of the growth factor means was below $0.07$, and the value of the variances of the initial status, asymptote slope, and distance between two intercepts was less than $0.17$. 

\begin{table}
\centering
\begin{threeparttable}
\setlength{\tabcolsep}{5pt}
\renewcommand{\arraystretch}{0.75}
\caption{Median and range of the Relative RMSE of Parameters in the Proposed Jenss-Bayley LCSMs ($10$ Repeated Measurements)}
\begin{tabular}{p{3.0cm}p{1cm}R{5.5cm}R{5.5cm}}
\hline
\hline
& \textbf{Para.} & \textbf{Reduced Model} & \textbf{Full Model} \\
\hline
& & Median (Range) & Median (Range) \\
\hline
\multirow{4}{*}{\textbf{\makecell[l]{Growth Factor \\ Means}}} 
& $\mu_{\eta_{0}}$ & $0.0047$ ($0.0036$, $0.0064$) & $0.0048$ ($0.0035$, $0.0064$) \\
& $\mu_{\eta_{1}}$ & $0.0370$ ($0.0191$, $0.0889$) & $0.0342$ ($0.0181$, $0.0698$) \\
& $\mu_{\eta_{2}}$ & $-0.0178$ ($-0.0288$, $-0.0107$) & $-0.0200$ ($-0.0300$, $-0.0100$) \\
& $\mu_{\gamma}$ & $-0.0158$ ($-0.0318$, $-0.0067$) & $-0.0131$ ($-0.0221$, $-0.0070$) \\
\hline
\multirow{4}{*}{\textbf{\makecell[l]{Growth Factor \\ Variances}}} 
& $\psi_{00}$ & $0.0944$ ($0.0643$, $0.1228$) & $0.0881$ ($0.0626$, $0.1165$) \\
& $\psi_{11}$ & $0.1676$ ($0.0651$, $0.4612$) & $0.1091$ ($0.0698$, $0.1691$) \\
& $\psi_{22}$ & $0.1757$ ($0.0702$, $0.3385$) & $0.1182$ ($0.0756$, $0.1639$) \\
& $\psi_{\gamma\gamma}$ & ---\tnote{a} & $0.5914$ ($0.1397$, NA)\tnote{b} \\
\hline
\hline
\end{tabular}
\label{tbl:rRMSE}
\begin{tablenotes}
\small
\item[a] {--- indicates that the relative RMSEs are not available from the reduced model.}\\
\item[b] {NA indicates that the bound of relative RMSE is not available. The model performance under the conditions with $0$ population value of the variance of the logarithmic ratio of the growth acceleration is of interest where the relative RMSE would go infinity. The median (range) of the RMSE of the logarithmic ratio of the growth acceleration for the proposed model is $0.0016$ ($0.0008$, $0.0036$).}
\end{tablenotes}
\end{threeparttable}
\end{table}

Table \ref{tbl:CP} presents the median and range of the CP of each parameter of interest for the proposed Jenss-Bayley LCSM and its reduced model. Overall, the full model performed well regarding empirical coverage as the median values of CPs of all parameters were near $0.95$, except for $\mu_{\eta_{2}}$ (i.e., the mean of the vertical distance between the two intercepts). One possible reason for the unsatisfied CP of $\mu_{\eta_{2}}$ is the underestimated SE. 

\begin{table}
\centering
\begin{threeparttable}
\setlength{\tabcolsep}{5pt}
\renewcommand{\arraystretch}{0.75}
\caption{Median and range of the Coverage Probabilities of Parameters in the Proposed Jenss-Bayley LCSMs ($10$ Repeated Measurements)}
\begin{tabular}{p{3.0cm}p{1cm}R{5.5cm}R{5.5cm}}
\hline
\hline
& \textbf{Para.} & \textbf{Reduced Model} & \textbf{Full Model} \\
\hline
& & Median (Range) & Median (Range) \\
\hline
\multirow{4}{*}{\textbf{\makecell[l]{Growth Factor \\ Means}}} 
& $\mu_{\eta_{0}}$ & $0.946$ ($0.9280$, $0.9580$) & $0.9506$ ($0.9233$, $0.9633$) \\
& $\mu_{\eta_{1}}$ & $0.939$ ($0.5540$, $0.9680$) & $0.9395$ ($0.7640$, $0.9690$) \\
& $\mu_{\eta_{2}}$ & $0.885$ ($0.3570$, $0.9690$) & $0.8025$ ($0.3398$, $0.9531$) \\
& $\mu_{\gamma}$ & $0.841$ ($0.1000$, $0.9570$) & $0.9380$ ($0.7720$, $0.9734$) \\
\hline
\multirow{4}{*}{\textbf{\makecell[l]{Growth Factor \\ Variances}}} 
& $\psi_{00}$ & $0.9235$ ($0.8280$, $0.9640$) & $0.9448$ ($0.9232$, $0.9664$) \\
& $\psi_{11}$ & $0.7885$ ($0.0010$, $0.9600$) & $0.9501$ ($0.9370$, $0.9723$) \\
& $\psi_{22}$ & $0.7475$ ($0.0290$, $0.9580$) & $0.9468$ ($0.8950$, $0.9660$) \\
& $\psi_{\gamma\gamma}$ & ---\tnote{a} & $0.9679$ ($0.9380$, $0.9910$) \\
\hline
\hline
\end{tabular}
\label{tbl:CP}
\begin{tablenotes}
\small
\item[a] {--- indicates that the coverage probabilities are not available from the reduced model.}
\end{tablenotes}
\end{threeparttable}
\end{table}

\subsection{Comparison between the Full and Reduced Jenss-Bayley LCSM}
\begin{figure}[!htbp]
\centering
\begin{subfigure}{.5\textwidth}
  \centering
  \includegraphics[width=1\linewidth]{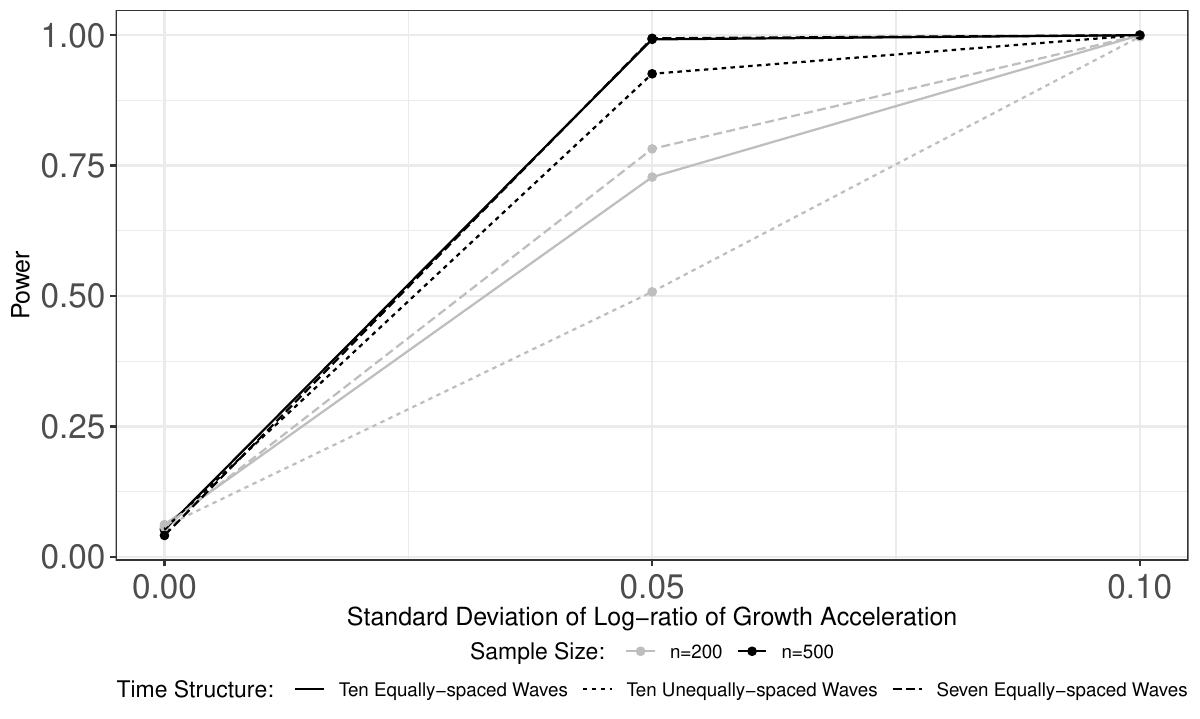}
  \caption{$\eta_{1i}\sim N(2.5, 1.0^2)$, $\theta_{\epsilon}=1$}
  \label{fig:PowerSres1}
\end{subfigure}%
\begin{subfigure}{.5\textwidth}
  \centering
  \includegraphics[width=1\linewidth]{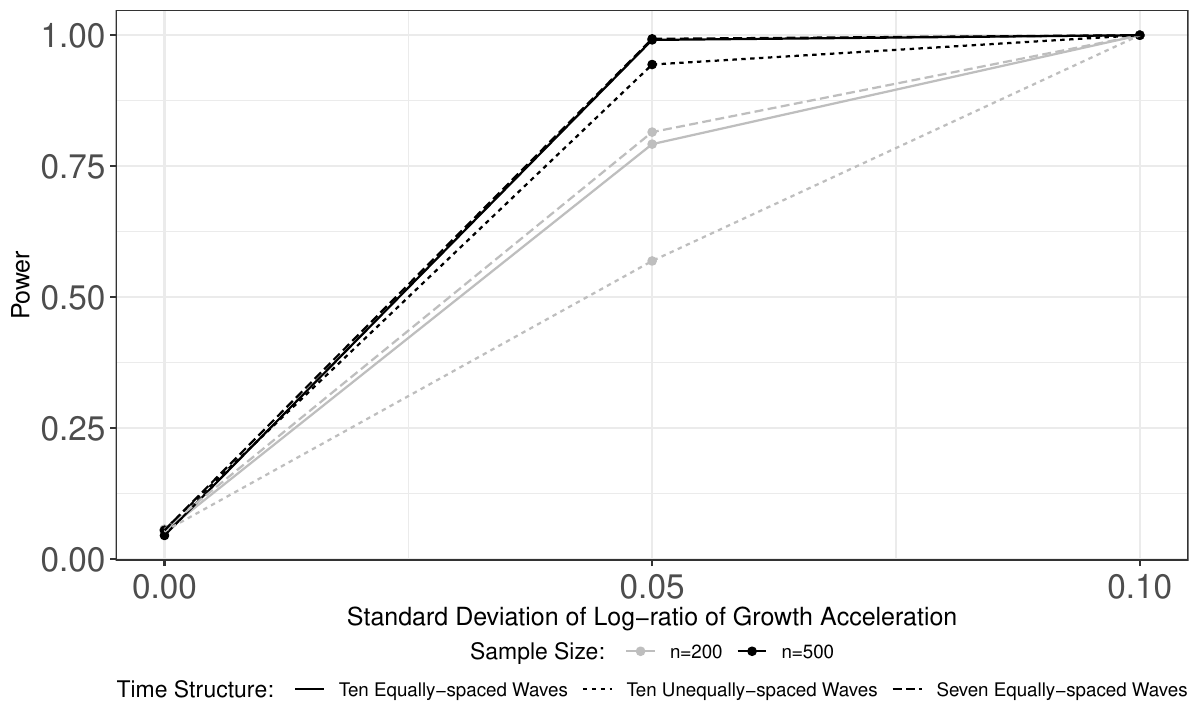}
  \caption{$\eta_{1i}\sim N(1.0, 0.4^2)$, $\theta_{\epsilon}=1$}
  \label{fig:PowerSres2}
\end{subfigure}
\vskip\baselineskip
\begin{subfigure}{.5\textwidth}
  \centering
  \includegraphics[width=1\linewidth]{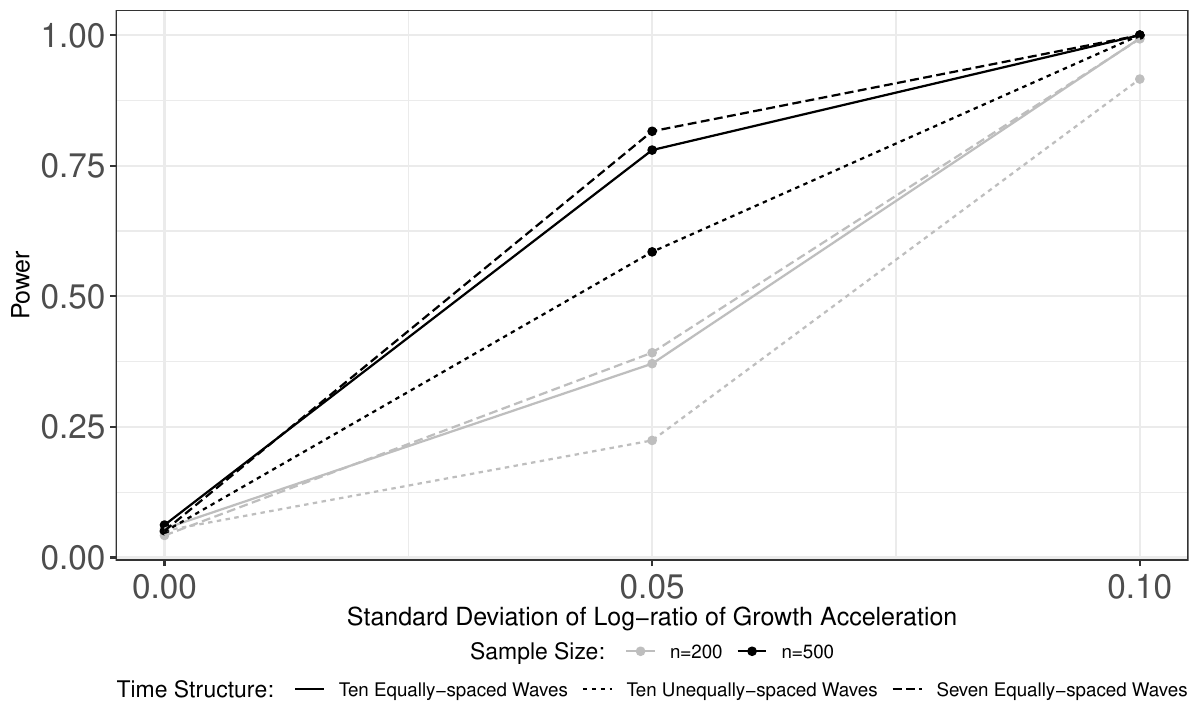}
  \caption{$\eta_{1i}\sim N(2.5, 1.0^2)$, $\theta_{\epsilon}=2$}
  \label{fig:PowerLres1}
\end{subfigure}%
\begin{subfigure}{.5\textwidth}
  \centering
  \includegraphics[width=1\linewidth]{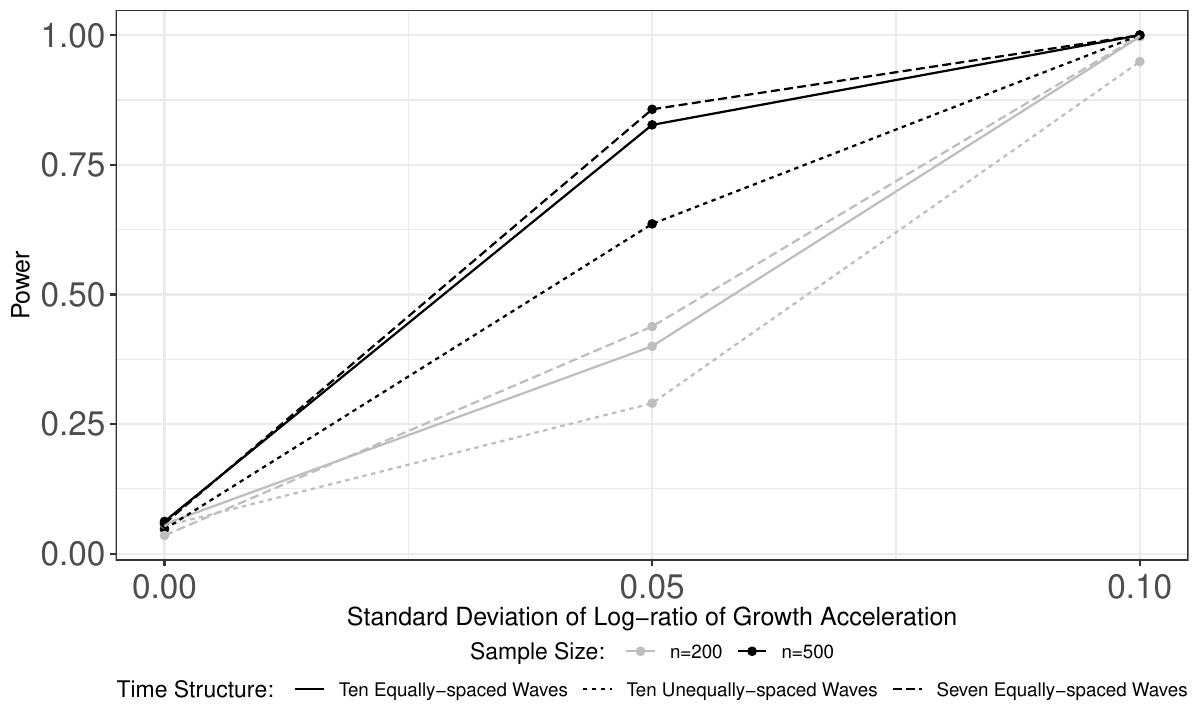}
  \caption{$\eta_{1i}\sim N(1.0, 0.4^2)$, $\theta_{\epsilon}=2$}
  \label{fig:PowerLres2}
\end{subfigure}
\caption{Statistical Power of Likelihood Ratio Test to Test Zero Variance of logarithmic ratio of the growth acceleration}
\label{fig:power}
\end{figure}

This section compares the proposed Jenss-Bayley latent change model with its reduced version through two perspectives. First, we summarize the factors that affect the statistical power to detect between-individual differences in the ratio of the growth acceleration. Second, we compare the two models regarding the performance metrics. Figure \ref{fig:power} describes the simulation result of the statistical power of the $4$ degree of freedom LRTs based on the $1000$ Monte Carlo replications of each condition. Each panel plots the power as a function of the varying ratio of the growth acceleration. It shows that the LRT controlled well for the Type I error rate since the size of the test of each condition without considering the ratio variance was around $0.05$. Additionally, greater values of the true between-individual differences in the ratio, more precise measurements, and larger sample size would improve the statistical power to detect the variance.

In terms of the performance metrics, the estimated variance of the linear asymptote and the distance between two intercepts from the reduced model were biased, although the relative biases of other parameters and the precision of each estimate from the two models are comparable, as shown in Tables \ref{tbl:rBias} and \ref{tbl:empSE}. In addition, the coverage probabilities generated by the reduced model were less satisfied, as shown in Table \ref{tbl:CP}.

\subsection{Comparison between the Proposed and Existing Jenss-Bayley LCSM}
We also compared the performance of the LCSM with the novel specification to that with the existing specification. We summarize the relative bias and empirical SE of each parameter from LCSMs with the existing specification in Tables \ref{tbl:rBias} and \ref{tbl:empSE}, respectively. From Table \ref{tbl:rBias}, we can see that the relative bias of the $\mu_{\eta_{2}}$ and $\psi_{22}$ could reach $0.46$ and $1.69$, respectively, which were much larger than the corresponding value from the LCSM with the novel specification. Additionally, the proposed LCSM improved the precision of the estimates for $\mu_{\eta_{2}}$ and $\psi_{22}$, as shown in Table \ref{tbl:empSE}. 

To summarize, based on our simulation study, the estimates from the proposed Jenss-Bayley LCSM were unbiased and precise, with the target coverage probabilities in general. Some factors, such as the number of repeated measurements and the placement of those measurements, influenced model performance. Specifically, more measurements, especially more measurements at an early stage, improved the model performance. This result aligns with the findings in existing studies, such as \citet{Timmons2015timing}. Through the simulation study, we found that the proposed Jenss-Bayley LCSM was robust under the conditions with the large standard deviation of the logarithmic ratio of the growth acceleration (i.e., $sd(\gamma)=0.10$), although it was less satisfactory when the standard deviation was small (i.e., $sd(\gamma)=0.05$). One possible explanation for this counter-intuitive finding is that we magnified the relative biases when replacing the negative estimates with zero if improper solutions occurred. 

\section{Application}\label{Sec:Application}
This section demonstrates how to employ the proposed model to estimate the individual ratio of the growth acceleration and obtain the individual instantaneous rate-of-change over time. This application has two goals. The first goal is to provide a set of feasible recommendations for real-world practices. Second, we want to understand how different modeling frameworks with the same function affect estimations;  therefore, we constructed a Jenss-Bayley LGCM with an individual ratio of the growth acceleration as a sensitivity analysis. We extracted $400$ students randomly from the Early Childhood Longitudinal Study Kindergarten Cohort: 2010-2011 (ECLS-K: 2011) with complete records of repeated reading item response theory (IRT) scaled scores and age at each wave\footnote{There are $n=18174$ participants in ECLS-K: 2011 is. After removing records with missing values (i.e., rows with any of NaN/-9/-8/-7/-1), we have $n=3418$ entries.}. 

ECLS-K: 2011 is a nationwide longitudinal study of US children enrolled in about $900$ kindergarten programs that started from the $2010-2011$ school year. In ECLS-K: 2011, children's reading ability was evaluated in nine waves: fall and spring of kindergarten ($2010-2011$), first grade ($2011-2012$) and second grade ($2012-2013$), respectively, as well as spring of $3^{rd}$ ($2014$), $4^{th}$ ($2015$) and $5^{th}$ ($2016$) grade, respectively. According to \citet{Le2011ECLS}, only about $30\%$ students were evaluated in the fall semester of $2011$ and $2012$. We used children's age (in years) at each wave to obtain individual measurement occasions in the analysis. In the subsample, $50.25\%$ and $49.75\%$ of children were boys and girls. Additionally, the selected sample was represented by $39.75\%$ White, $7.25\%$ Black, $41.75\%$ Latinx, $5.75\%$ Asian, and $5.50\%$ others. We provide the raw trajectories of $100$ randomly selected individuals and the smooth line in Figure \ref{fig:raw}. We can see that the development in reading ability was steep at the early stage and then slowed down, and therefore, the Jenss-Bayley functional form is one candidate to describe the underlying change pattern.

\subsection{Main Analysis}\label{A:main}
In this section, we fit the full and reduced Jenss-Bayley LCSMs in the framework of individual measurement occasions. Table \ref{tbl:info} lists the estimated likelihood, information criteria, including the Akaike information criterion (AIC) and Bayesian Information Criteria (BIC), residuals, and the number of parameters of each LCSM. As shown in Table \ref{tbl:info}, the full Jenss-Bayley LCSM has a greater estimated likelihood, a smaller AIC, a smaller BIC, and smaller residual variance. In addition, the p-value of the LRT to test the variability of the growth acceleration ratio was $<0.0001$. All information led to the unequivocal selection of the full Jenss-Bayley LCSM. 

\begin{table}
\centering
\resizebox{1.0\textwidth}{!}{
\begin{threeparttable}
\small
\setlength{\tabcolsep}{4pt}
\renewcommand{\arraystretch}{0.6}
\caption{Summary of Model Fit Information For the Models}
\begin{tabular}{lrrrrr}
\hline
\hline
\multicolumn{6}{c}{\textbf{Proposed Latent Change Score Models}} \\
\hline
\textbf{Model} & \textbf{-2ll} & \textbf{AIC}  & \textbf{BIC}  & \textbf{\# of Para.} & \textbf{Residual}  \\
\hline
Full Jenss-Bayley Latent Change Score Model & $26105$ & $26135$ & $26195$ & $15$ & $41.43$ \\
\hline
Reduced Jenss-Bayley Latent Change Score Model & $26248$ & $26270$ & $26314$ & $11$ & $44.31$ \\
\hline
\hline
\multicolumn{6}{c}{\textbf{Latent Growth Curve Models}} \\
\hline
\textbf{Model} & \textbf{-2ll} & \textbf{AIC}  & \textbf{BIC}  & \textbf{\# of Para.} & \textbf{Residual}  \\
\hline
Full Jenss-Bayley Latent Growth Curve Model & $26118$ & $26148$ & $26207$ & $15$ & $42.02$ \\
\hline
Reduced Jenss-Bayley Latent Growth Curve Model & $26225$ & $26247$ & $26291$ & $11$ & $44.68$ \\
\hline
\hline
\end{tabular}
\label{tbl:info}
\end{threeparttable}}
\end{table}

Table \ref{tbl:est} presents the estimates of the parameters of interest. The development in reading skills slowed down gradually as the logarithmic ratio of the growth acceleration was negative. On average,  the ratio of the growth acceleration at any given year to the acceleration at the preceding year was $0.70$ (i.e., $\exp(-0.35)$), indicating that the development of reading ability was a decelerating process from Grade K to Grade $5$. The estimated variance of the logarithmic ratio of the growth acceleration was $0.18$. It tells us that individual students had `personal' ratios of the growth accelerations for reading ability. About $68\%$ of the time, the ratio was in the range of ($0.46$, $1.08$) (i.e., ($\exp(-0.35-\sqrt{0.18})$, $\exp(-0.35+\sqrt{0.18})$)), and about $95\%$ of the time, the ratio was in the range of ($0.30$, $1.65$) (i.e., ($\exp(-0.35-\sqrt{0.18}\times2)$, $\exp(-0.35+\sqrt{0.18}\times2)$)). On average, the linear asymptote slope is $0.36$ per year, which is not statistically significant. It suggests that the development of reading ability reached a stable status. Additionally, the linear asymptote slope and the vertical distance between two intercepts varied substantially among individuals suggested by their large and statistically significant variances.

\begin{table}
\centering
\resizebox{1.0\textwidth}{!}{
\begin{threeparttable}
\small
\renewcommand{\arraystretch}{0.6}
\caption{Estimates of the Jenss-Bayley Latent Change Score Model with Individual Ratio of the growth acceleration}
\begin{tabular}{lR{2cm}R{2cm}R{2cm}|R{2cm}R{2cm}R{2cm}}
\hline
\hline
\textbf{Growth Factor} & \multicolumn{3}{c}{\textbf{Initial Status}\tnote{a}} & \multicolumn{3}{c}{\textbf{Slope of Linear Asymptote}} \\
\hline
\textbf{Parameter} & \textbf{Estimate} & \textbf{SE} & \textbf{P value} & \textbf{Estimate} & \textbf{SE} & \textbf{P value} \\
\hline
\textbf{Mean} & $54.35$ & $0.63$ & $<0.0001^{\ast}$ & $0.36$ & $1.28$ &$0.7774$ \\
\textbf{Variance} & $113.44$ &$10.36$ & $<0.0001^{\ast}$ & $243.01$ & $95.26$ & $0.0107^{\ast}$ \\
\hline
\hline
\textbf{Growth Factor} & \multicolumn{3}{c}{\textbf{Vertical Distance between Two Intercepts}} & \multicolumn{3}{c}{\textbf{logarithmic ratio of the growth acceleration}}\\
\hline
\textbf{Parameter} & \textbf{Estimate} & \textbf{SE} & \textbf{P value} & \textbf{Estimate} & \textbf{SE} & \textbf{P value} \\
\hline
\textbf{Mean} & $-113.88$ & $11.54$ & $<0.0001^{\ast}$ &$-0.35$ & $0.03$ &$<0.0001^{\ast}$ \\
\textbf{Variance} & $23199.78$ &$10319.02$ & $0.0246^{\ast}$ & $0.18$ & $0.05$ & $0.0002^{\ast}$ \\
\hline
\hline
\end{tabular}
\label{tbl:est}
\begin{tablenotes}
\small
\item[a] For this analysis, the initial status of reading ability is the reading ability at five years old. \\
\item[b] $^{\ast}$ indicates statistical significance at $0.05$ level.
\end{tablenotes}
\end{threeparttable}}
\end{table}

\begin{figure}[!htbp]
\centering
  \includegraphics[width=1.0\linewidth]{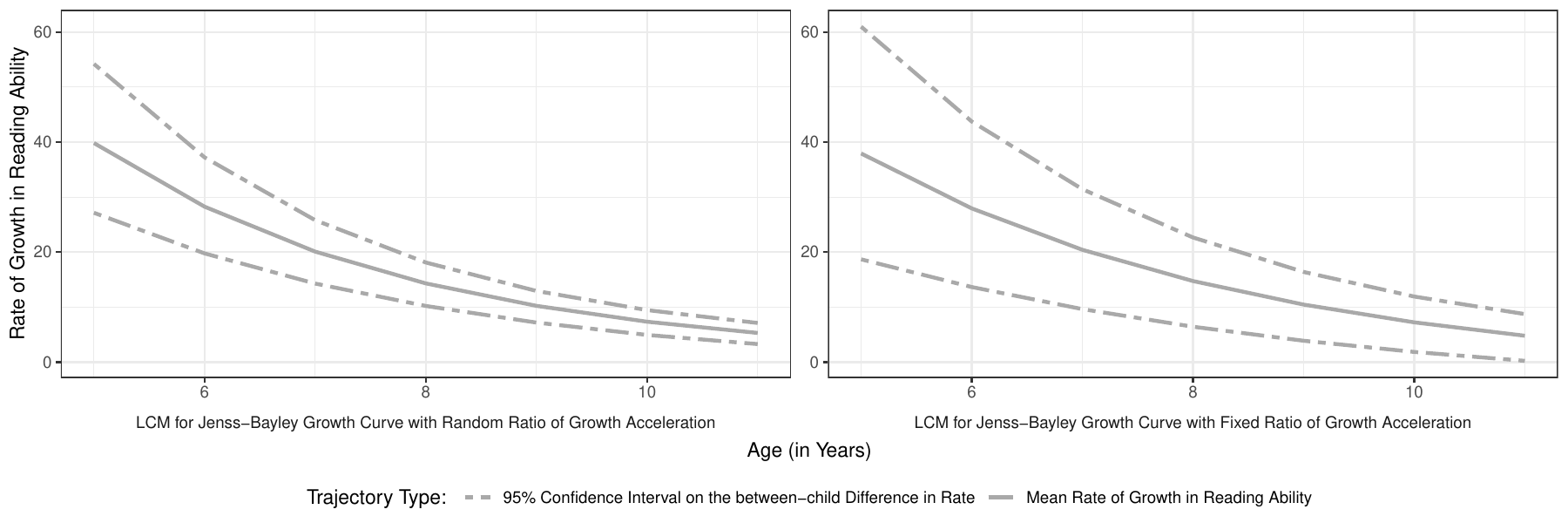}
  \caption{Longitudinal Plot of the Mean Yearly Growth Rate from Jenss-Bayley LCSMs}
\label{fig:rate_mean}
\end{figure}

\begin{figure}[!htbp]
\centering
  \includegraphics[width=1.0\linewidth]{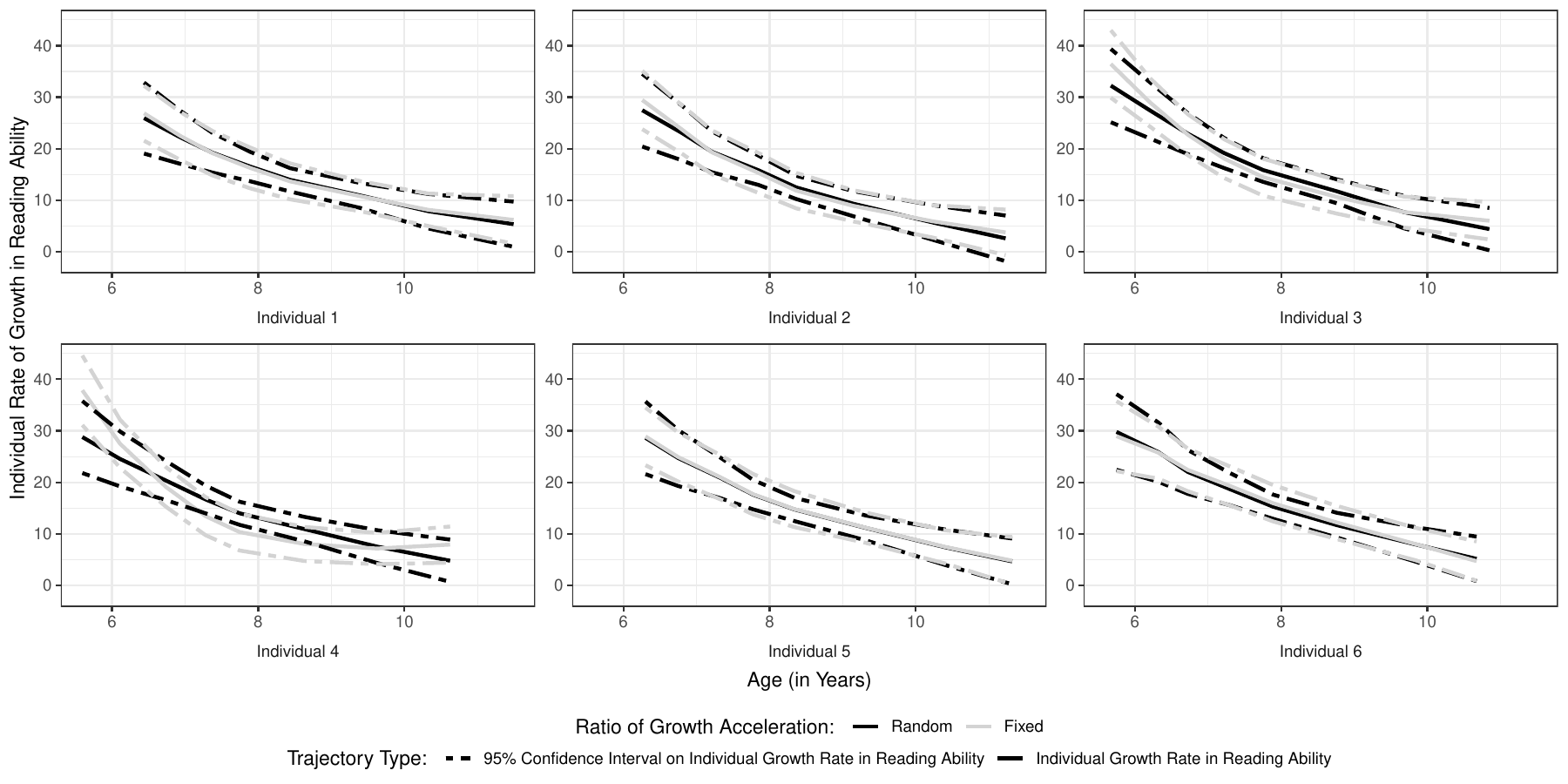}
  \caption{Longitudinal Plot of the Individual Yearly Growth Rate from Jenss-Bayley LCSMs}
\label{fig:rate_ind}
\end{figure}

To further understand how the individual ratio of the growth acceleration affects the rate-of-change over time, we provide the mean values and individual scores of yearly rate-of-change over time obtained by the full and the reduced Jenss-Bayley LCSM in Figures \ref{fig:rate_mean} and \ref{fig:rate_ind}, respectively. As shown in Figure \ref{fig:rate_mean}, the yearly rate-of-change estimated from both full and reduced models was expected to slow down in the late stage of the study, as did the magnitude of between-individual differences. However, the $95\%$ confidence interval on the rate-of-change of the full model was narrower than that from the reduced model. Figure \ref{fig:rate_ind} plots the individual yearly rate-of-change for six individuals from the data set. It shows that the $r-t$ graphs of the two models were similar for most individuals. However, for the third and fourth individuals, the estimated growth rate from the full model was more gradual than that from the reduced model. We then examined the observed growth trajectory (see Figure \ref{fig:growth_ind}) for each individual and found that one possible reason for the steep rate estimated by the reduced model for Individuals $3$ and $4$ is the fluctuation in academic performance. These findings suggest that the Jenss-Bayley LCSM with an individual ratio of the growth acceleration can capture the growth rate more precisely than the model with a fixed ratio in this analysis. 

Another important output of the LCSM is the estimates of the change-from-baseline at each post-baseline time point, which is a common metric to examine the change in longitudinal data analyses. In Figure \ref{fig:CHG}, we plot the model-implied change-from-baseline on the smooth line of the corresponding observed values for the development of reading ability. The figure shows that both the proposed LCSM and its reduced version can estimate the amount of change-from-baseline satisfactorily.

\subsection{Sensitivity Analysis}
\begin{figure}[!htbp]
\centering
  \includegraphics[width=1.0\linewidth]{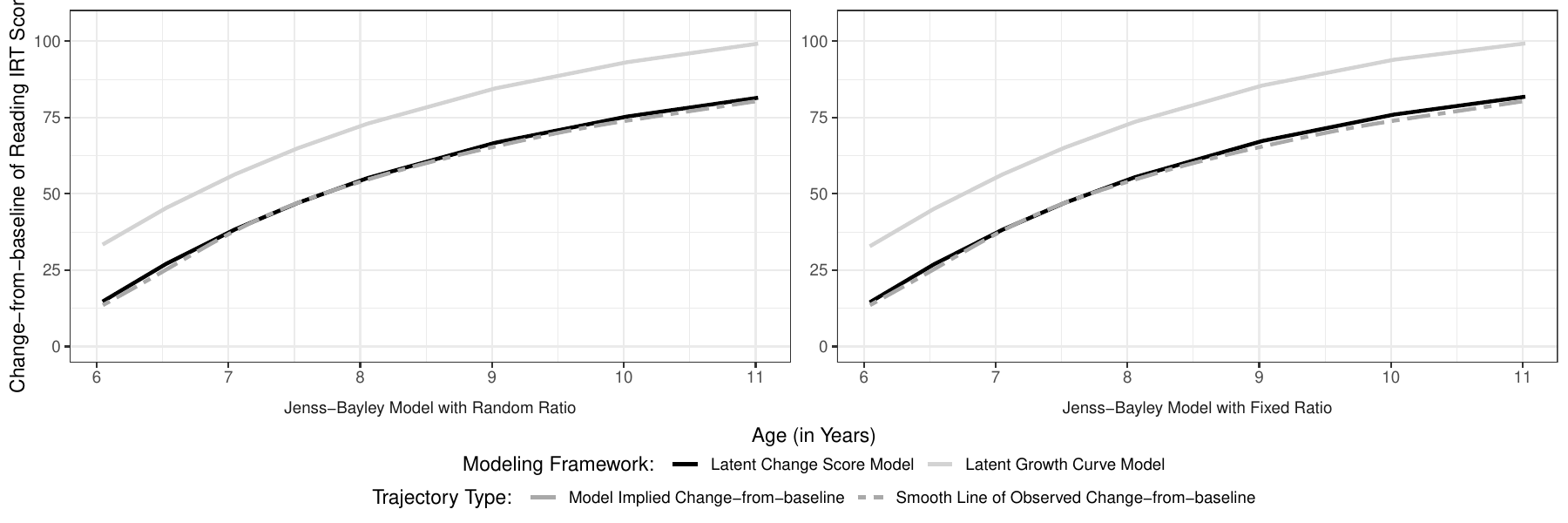}
  \caption{Model-implied Change-from-baseline and Smooth Line of Observed Change-from-baseline}
\label{fig:CHG}
\end{figure}

We then constructed the Jenss-Bayley LGCMs as a sensitivity analysis. We list the estimated likelihood, AIC, BIC, and residual variance in Table \ref{tbl:info}. From the table, we can see that the full Jenss-Bayley LGCM  outperformed its reduced version. We also derived the values of change-from-baseline for the Jenss-Bayley LGCMs and provided the plots in Figure \ref{fig:CHG}. It can be seen that the Jenss-Bayley LGCMs tended to overestimate the change-from-baseline. One possible reason for the poor performance of the  Jenss-Bayley LGCMs in evaluating the change is that they underestimated the intercept means (the estimated mean value of the intercept of the Jenss-Bayley LGCM with a random ratio of the growth acceleration and its reduced model was $36.63$ and $35.44$, respectively).

\section{Discussion}\label{sec:Discussion}
This article extends an existing Jenss-Bayley LCSM to estimate an individual ratio of the growth acceleration in the framework of individual measurement occasions. We approximate the latent change score using the product of the instantaneous growth rate at the mid-point of consecutive measurement occasions and the time interval between the two occasions to address multiple challenges of the implementation of the existing Jenss-Bayley LCSM. More importantly, we employ the Taylor series expansion to address a nonlinear relationship between a target function and a random coefficient and allow for an individual ratio of the growth acceleration. 

We examine the proposed model by the simulation study on the Jenss-Bayley LGCM-implied data structure to investigate whether and how the approximation affects the model performance. We compare the proposed model to its reduced model and show that the LRT controls well for the Type I error rate and can detect between-individual differences in the ratio of the growth acceleration. Therefore, other than the AIC and BIC, we can also utilize the LRT to decide the model preference in practice. We demonstrate how to implement the proposed model on a subset with $n=400$ from ECLS-K: $2011$. 

\subsection{Practical Considerations}\label{D:practical}
This section provides a set of recommendations for empirical researchers based on the simulation study and the real-world data analysis. First, the Jenss-Bayley model, determined by four parameters, can be viewed as a combination of an exponential and linear growth model. It has a steep initial development followed by a level-off growth. Accordingly, we recommend visualizing the raw trajectories to check whether they demonstrate such patterns as we did in the Application section. Second, for an empirical study, we still recommend assessing the issue of improper solutions. The simulation study showed that almost all improper solutions were observed when we over-specified the model. Based on this result, an improper variance or correlation may suggest that the ratio of the growth acceleration is roughly similar across all individuals. Third, based on the output of the simulation study, the proposed Jenss-Bayley LCSM performed well generally. However, under challenging conditions such as a large standard deviation of the logarithmic ratio of the growth acceleration\footnote{By `large standard deviation of the logarithmic ratio of the growth acceleration', we mean more challenging conditions than what we considered in the simulation design, for example, $sd(\gamma)=0.20$.}, the estimates of the variance of the vertical distance between two intercepts exhibited some bias greater than $10\%$. It suggests that the estimate of $\psi_{22}$ is potentially over-estimated and unreliable in a real-world data analysis. 

Additionally, the time unit selection affects the estimates of the ratio of the growth acceleration since it measures the ratio of the growth acceleration at two consecutive time points and changes with the time unit. It suggests that the `personal' ratio of the growth acceleration may not be detectable if a small unit is employed. For example, the ratio of the growth acceleration in reading development only has a fixed effect if we use age-in-month instead of age-in-year in the case that we demonstrated in the Application section. 

Another advantage of the Jenss-Bayley LCSM over its LGCM is that the LCSM can provide more reliable estimated values of change-from-baseline, as shown in the sensitivity analysis. This better performance of the LCSM lies in that we do not utilize the pre-specified functional form to capture the change patterns; instead, we employ the first derivative of the function to constrain the pattern of rate-of-change, which is unrelated to the initial status (see Figure \ref{fig:path_novel}). To maximize the likelihood function, the LCSM tends to converge to a solution with an optimized initial status and the first derivative of the trajectory function, while the LGCM tends to fit the whole trajectory. 

\subsection{Methodological Considerations and Future Directions}\label{D:method}
There are multiple directions for future exploration. First, the proposed expression for the latent change score can be generalized to LCSMs with other functional forms. We also provide the code of the quadratic and exponential LCSM with the novel expression of change score on the Github website for researchers who are interested in using them. The proposed expression for latent change scores can also be extended to other commonly used LCSMs, such as proportional change models and dual change models. Additionally, we can extend the proposed Jenss-Bayley LCSM to the dual change modeling framework by replacing the time-invariant additive constant with a Jenss-Bayley functional form to investigate more complicated change patterns as recommended by \citet{Hamagami2018LCSM}. Moreover, the proposed model can also be extended to investigate the covariates to explain the individual differences in the rate-of-change.

In a pilot simulation study, we noticed that the proposed model generated biased estimates for the random effect of the vertical distance between two intercepts when the standard deviation of the logarithmic ratio of the growth acceleration was set as $0.2$. Other than the approximation introduced by the Taylor series expansion, there are additional possible explanations. First, $\gamma_{i}$ could achieve $-0.3$ when its distribution follows $\text{N}(-0.7, 0.2^2)$, and a growth curve with this less negative ratio of the growth acceleration approaches to the linear asymptote relatively late. The study duration we considered in the simulation design may not be sufficiently long to capture its entire change pattern. It suggests that it is important to determine the study duration based on available growth factors when designing a longitudinal study where the Jenss-Bayley model is a candidate for the underlying change pattern. Another possible reason for the bias is that we constructed a LCSM on a LGCM-implied data structure. Additionally, as shown in the Application section, the estimates from the LCSM and those from the LGCM could be different even if we specified the same functional form for the underlying change patterns. The comparison between the LCSM and the corresponding LGCM by a simulation study is beyond the scope of the current project, but it can be examined in future work. 

\subsection{Concluding Remarks}\label{D:conclude}
This article demonstrates a novel expression for the latent change score in the Jenss-Bayley LCSM to allow (1) unequally-spaced waves and (2) individually varying measurement occasions around each wave. We also demonstrate that the first-order Taylor series expansion, one popular linearization approach, can be used to estimate an individual growth acceleration ratio. The results of the simulation study and the real-world data analysis demonstrate the model's valuable capabilities of estimating the ratio of the growth acceleration and its variance in the framework of individual measurement occasions. As discussed above, the proposed method can be generalized in practice and further examined in methodology.

\bibliographystyle{apalike}
\bibliography{Extension5}

\appendix
\renewcommand{\theequation}{A.\arabic{equation}}
\setcounter{equation}{0}

\renewcommand{\thesection}{Appendix \Alph{section}}
\renewcommand{\thesubsection}{A.\arabic{subsection}}

\section{Formula Derivation}\label{Supp:1}
\subsection{Taylor Series Expansion}\label{Supp:1a}
For the $i^{th}$ individual, suppose we define a function $f(\gamma_{i})$ and its first derivative with respect to $\gamma_{i}$, shown below
\begin{equation}\nonumber
f(\gamma_{i})=dy_{ij}=\eta_{1i}+\eta_{2i}\gamma_{i}(\exp(\gamma_{i}t_{ij}))
\end{equation}
and
\begin{equation}\nonumber
f^{'}(\gamma_{i})=\eta_{2i}(\exp(\gamma_{i}t_{ij}))+\eta_{2i}\gamma_{i}t_{ij}(\exp(\gamma_{i}t_{ij})),
\end{equation}
respectively. Then the Taylor series expansion of $f(\gamma_{i})$ can be expressed as
\begin{equation}\nonumber
\begin{aligned}
f(\gamma_{i})&=f(\mu_{\gamma})+\frac{f'(\mu_{\gamma})}{1!}(\gamma_{i}-\mu_{\gamma})+\cdots\\
&=\eta_{1i}+\eta_{2i}\mu_{\gamma}(\exp(\mu_{\gamma}t_{ij}))+(\gamma_{i}-\mu_{\gamma})\bigg[\eta_{2i}\exp(\mu_{\gamma}t_{ij})+\eta_{2i}\mu_{\gamma}t_{ij}\exp(\mu_{\gamma}t_{ij})\bigg]\\
&\approx \eta_{1i}+\eta_{2i}\mu_{\gamma}(\exp(\mu_{\gamma}t_{ij}))+(\gamma_{i}-\mu_{\gamma})\bigg[\mu_{\eta_{2}}\exp(\mu_{\gamma}t_{ij})(1+\mu_{\gamma}t_{ij})\bigg],
\end{aligned}
\end{equation}
from which we then have the reparameterized growth factors and the corresponding factor loadings for the $i^{th}$ individual.

\subsection{Deviation of the Matrix of Factor Loadings for the Proposed Model}\label{Supp:1b}
We first write the derivative specified in Equation \ref{eq:LCSM4} as the matrix form
\begin{equation}\label{eq:LCSM4_matrix}
\boldsymbol{dy}_{i}\approx\boldsymbol{\Lambda}_{ri}\times\boldsymbol{\eta}_{ri},
\end{equation}
where $\boldsymbol{dy}_{i}$ is a $(J-1)\times1$ vector of the instantaneous rate-of-change midway through each time interval between two consecutive measurement occasions for the $i^{th}$ individual (in which $J$ is the number of measurements), $\boldsymbol{\eta}_{ri}$ is a $3\times1$ vector of growth factors related to the growth rate for the individual $i$ (i.e., $\eta_{1i}$, $\eta_{2i}$ and $\gamma_{i}-\mu_{\gamma}$), and $\boldsymbol{\Lambda}_{ri}$ is a $(J-1)\times3$ matrix of the corresponding factor loadings. The notation `$r$' in the subscript in Equation \ref{eq:LCSM4_matrix} indicates that the growth factors and the corresponding factor loadings are related to the rate-of-change.

For the $i^{th}$ individual, the rate-related growth factors and the corresponding factor loadings are
\begin{equation}\nonumber
\boldsymbol{\eta}_{ri}=\left(\begin{array}{rrr}
\eta_{1i} & \eta_{2i} & \gamma_{i}-\mu_{\gamma}
\end{array}\right)^{T}
\end{equation}
and
\begin{equation}\nonumber
\begin{aligned}
&\boldsymbol{\Lambda}_{ri} = \left(\begin{array}{rrr}
1 & \mu_{\gamma}\exp(\mu_{\gamma}t_{ij\_\text{mid}}) & \mu_{\eta_{2}}\exp(\mu_{\gamma}t_{ij\_\text{mid}})(1+\mu_{\gamma}t_{ij\_\text{mid}})
\end{array}\right)
&(j=2,\cdots, J),
\end{aligned}
\end{equation}
respectively. Note that the first, second, and third column of $\boldsymbol{\Lambda}_{ri}$ in the above equation represents the linear asymptote slope, the exponential slope, and the `additional' slope related to the fourth growth factor, respectively. We then define a $J\times(J-1)$ matrix $\Omega_{i}$ for each individual to represent the `definition variables'
\begin{equation}\label{eq:omega}
 \boldsymbol{\Omega}_{i}=\begin{pmatrix}
  0 & 0 & \cdots & \cdots & \cdots & 0 \\
  t_{i2}-t_{i1} & 0 & 0 & \cdots & \cdots & 0 \\
  t_{i2}-t_{i1} & t_{i3}-t_{i2} & 0 & 0 & \cdots & 0 \\
  \cdots & \cdots & \cdots & \cdots & \cdots & \cdots \\
  t_{i2}-t_{i1} & t_{i3}-t_{i2} & t_{i4}-t_{i3} & \cdots & \cdots & t_{ij}-t_{i(j-1)} \\
\end{pmatrix}.
\end{equation}
The elements in the first row of $\Omega_{i}$ are all zero because there is no cumulation in time $t$ when $j=1$. Similarly, we only have one non-zero element in the second row because we have only one interval for each individual, $t_{i2}-t_{i1}$, at $j=2$, we have two intervals for each individual, $t_{i2}-t_{i1}$ and $t_{i3}-t_{i2}$, at $j=3$ (i.e., the third row), and so forth. With this definition, the product of $\boldsymbol{\Omega}_{i}$ and $\boldsymbol{\Lambda}_{ri}$ is the grey part of $\boldsymbol{\Lambda}_{i}$, of which the $j^{th}$ row can be interpreted as the cumulative amount of each slope (i.e., each element in $\boldsymbol{\Lambda}_{ri}$) until time $t_{j}$.

\subsection{Individual Scores of Latent Variables}\label{Supp:1c}
For the $i^{th}$ individual, the joint distribution of repeated measurements $\boldsymbol{y}_{i}$ and all latent variables $\boldsymbol{\eta}_{ai}$ (i.e., $\boldsymbol{\eta}_{ai}=(\begin{array}{ccc}\boldsymbol{\eta}^{T}_{i} & \boldsymbol{ly}^{T}_{i} & \boldsymbol{dy}^{T}_{i}\end{array})^{T}$) is
\begin{equation}\nonumber
\begin{pmatrix}
\boldsymbol{y}_{i} \\ \boldsymbol{\eta}_{ai} 
\end{pmatrix}\sim\text{MVN}\bigg(\begin{pmatrix}
\boldsymbol{\mu}_{i} \\ \boldsymbol{\mu}_{\boldsymbol{\eta}i}
\end{pmatrix}, \begin{pmatrix}
\boldsymbol{\Lambda}_{ai}\boldsymbol{\Psi}_{\boldsymbol{\eta}i}\boldsymbol{\Lambda}^{T}_{ai} + \theta_{\epsilon}\boldsymbol{I} & \boldsymbol{\Lambda}_{ai}\boldsymbol{\Psi}_{\boldsymbol{\eta}i} \\
\boldsymbol{\Psi}_{\boldsymbol{\eta}i}\boldsymbol{\Lambda}^{T}_{ai} & \boldsymbol{\Psi}_{\boldsymbol{\eta}i} 
\end{pmatrix}
\bigg).
\end{equation}
In this section, we provide the expression for the vectors and matrices that appeared in the distribution. Corresponding to $\boldsymbol{\eta}_{ai}$, $\boldsymbol{\mu}_{\boldsymbol{\eta}_i}$ is defined as $\boldsymbol{\mu}_{\boldsymbol{\eta}_i}=(\begin{array}{ccc}\boldsymbol{\mu}^{T}_{\boldsymbol{\eta}} & \boldsymbol{\mu}^{T}_{\boldsymbol{dy}i} & \boldsymbol{\mu}^{T}_{\boldsymbol{ly}i}\end{array})^{T}$. We can calculate $\boldsymbol{\mu}_{\boldsymbol{dy}i}$ and $\boldsymbol{\mu}_{\boldsymbol{ly}i}$ 
\begin{equation}\nonumber
\boldsymbol{\mu}_{\boldsymbol{dy}i}=\boldsymbol{\Lambda}_{ri}\times\boldsymbol{\mu}_{\boldsymbol{\eta}_{r}},
\end{equation}
and
\begin{equation}\nonumber
\boldsymbol{\mu}_{\boldsymbol{ly}i}=\boldsymbol{\Lambda}_{i}\times\boldsymbol{\mu}_{\boldsymbol{\eta}},
\end{equation}
respectively, where $\boldsymbol{\mu}_{\boldsymbol{\eta}_{r}}$ is the mean vector of the growth factors related to the rate-of-change. For the $i^{th}$ individual, the variance-covariance matrix of all latent variables $\boldsymbol{\Psi}_{\boldsymbol{\eta}i}$ can be expressed as
\begin{equation}
\boldsymbol{\Psi}_{\boldsymbol{\eta}i}=\begin{pmatrix}\nonumber
\boldsymbol{\Psi}_{\boldsymbol{\eta}} & \boldsymbol{0}_{4\times J} & \boldsymbol{0}_{4\times(J-1)} \\ \boldsymbol{0}_{J\times 4} & \boldsymbol{0}_{J\times J} & \boldsymbol{0}_{J\times (J-1)} \\
\boldsymbol{0}_{(J-1)\times 4} & \boldsymbol{0}_{(J-1)\times J} & \boldsymbol{0}_{(J-1)\times(J-1)}
\end{pmatrix}.
\end{equation}
Since $\boldsymbol{ly}_{i}$ and $\boldsymbol{dy}_{i}$ are not freely estimated in the model specification, we put $0$'s in the block matrices that are for the variance-covariance structures of $\boldsymbol{ly}_{i}$ and $\boldsymbol{dy}_{i}$ as well as their covariances with other blocks. 

Additionally, the matrix $\boldsymbol{\Lambda}_{ai}$ specifies the relationship between the latent variables $\boldsymbol{\eta}_{ai}$ and the repeated outcome $\boldsymbol{y}_{i}$, which can be expressed as 
\begin{equation}\nonumber
\boldsymbol{\Lambda}_{ai}=\begin{pmatrix} \boldsymbol{\Lambda}_{i} & \boldsymbol{0}_{J\times J} & \boldsymbol{0}_{J\times(J-1)}
\end{pmatrix}.
\end{equation}

\section{More Results}\label{Supp:2}
\begin{table}[!htbp]
\centering
\begin{threeparttable}
\setlength{\tabcolsep}{4pt}
\renewcommand{\arraystretch}{0.72}
\caption{Median and range of Performance Measures of Parameters in the Proposed Jenss-Bayley LCSMs ($7$ Repeated Measurements)}
\begin{tabular}{p{3.0cm}p{1cm}R{5.5cm}R{5.5cm}}
\hline
\hline
\multicolumn{4}{c}{\textbf{Relative Bias}} \\
\hline
\hline
& \textbf{Para.} & \textbf{Reduced Model} & \textbf{Full Model} \\
\hline
& & Median (Range) & Median (Range) \\
\hline
\multirow{4}{*}{\textbf{\makecell[l]{Growth Factor \\ Means}}} 
& $\mu_{\eta_{0}}$ & $-0.0002$ ($-0.0005$, $0.0002$) & $0.0000$ ($-0.0005$, $0.0004$) \\
& $\mu_{\eta_{1}}$ & $0.0062$ ($-0.0049$, $0.0516$) & $0.0011$ ($-0.0059$, $0.0256$) \\
& $\mu_{\eta_{2}}$ & $0.0450$ ($0.0341$, $0.0496$) & $0.0472$ ($0.0396$, $0.0496$) \\
& $\mu_{\gamma}$ & $0.0063$ ($-0.0038$, $0.0214$) & $0.0011$ ($-0.0037$, $0.0070$) \\
\hline
\multirow{4}{*}{\textbf{\makecell[l]{Growth Factor \\ Variances}}} 
& $\psi_{00}$ & $-0.0154$ ($-0.0357$, $0.0001$) & $-0.0027$ ($-0.0090$, $0.0028$) \\
& $\psi_{11}$ & $0.1229$ ($-0.0097$, $0.4056$) & $0.0112$ ($-0.0005$, $0.0678$) \\
& $\psi_{22}$ & $0.2252$ ($0.0914$, $0.4033$) & $0.1340$ ($0.1004$, $0.1527$) \\
& $\psi_{\gamma\gamma}$ & ---\tnote{a} & $0.4657$ ($-0.0360$, NA)\tnote{b} \\
\hline
\hline
\multicolumn{4}{c}{\textbf{Empirical Standard Errors}} \\
\hline
\hline
& \textbf{Para.} & \textbf{Reduced Model} & \textbf{Full Model} \\
\hline
& & Median (Range) & Median (Range) \\
\hline
\multirow{4}{*}{\textbf{\makecell[l]{Growth Factor \\ Means}}} 
& $\mu_{\eta_{0}}$ & $0.2392$ ($0.1825$, $0.3124$) & $0.2349$ ($0.1800$, $0.3149$) \\
& $\mu_{\eta_{1}}$ & $0.0529$ ($0.0341$, $0.0831$) & $0.0526$ ($0.0340$, $0.0854$) \\
& $\mu_{\eta_{2}}$ & $0.4302$ ($0.3156$, $0.5549$) & $0.4161$ ($0.3143$, $0.5570$) \\
& $\mu_{\gamma}$ & $0.0097$ ($0.0060$, $0.0161$) & $0.0098$ ($0.0060$, $0.0160$) \\
\hline
\multirow{4}{*}{\textbf{\makecell[l]{Growth Factor \\ Variances}}} 
& $\psi_{00}$ & $1.3900$ ($1.0527$, $1.8605$) & $1.4079$ ($1.0789$, $1.8801$) \\
& $\psi_{11}$ & $0.0659$ ($0.0268$, $0.1346$) & $0.0749$ ($0.0357$, $0.1386$) \\
& $\psi_{22}$ & $3.8992$ ($2.7568$, $5.6019$) & $4.1531$ ($3.0468$, $5.9793$) \\
& $\psi_{\gamma\gamma}$ & --- & $0.0021$ ($0.0010$, $0.0046$) \\
\hline
\hline
\multicolumn{4}{c}{\textbf{Relative RMSE}} \\
\hline
\hline
& \textbf{Para.} & \textbf{Reduced Model} & \textbf{Full Model} \\
\hline
& & Median (Range) & Median (Range) \\
\hline
\multirow{4}{*}{\textbf{\makecell[l]{Growth Factor \\ Means}}} 
& $\mu_{\eta_{0}}$ & $0.0048$ ($0.0037$, $0.0063$) & $0.0047$ ($0.0036$, $0.0063$) \\
& $\mu_{\eta_{1}}$ & $0.0370$ ($0.0199$, $0.0785$) & $0.0341$ ($0.0194$, $0.0676$) \\
& $\mu_{\eta_{2}}$ & $-0.0471$ ($-0.0527$, $-0.0357$) & $-0.0492$ ($-0.0528$, $-0.0411$) \\
& $\mu_{\gamma}$ & $-0.0180$ ($-0.0309$, $-0.0091$) & $-0.0144$ ($-0.0239$, $-0.0092$) \\
\hline
\multirow{4}{*}{\textbf{\makecell[l]{Growth Factor \\ Variances}}} 
& $\psi_{00}$ & $0.0901$ ($0.0683$, $0.1174$) & $0.0880$ ($0.0676$, $0.1175$) \\
& $\psi_{11}$ & $0.1645$ ($0.0663$, $0.4353$) & $0.1225$ ($0.0738$, $0.1947$) \\
& $\psi_{22}$ & $0.2534$ ($0.1241$, $0.4300$) & $0.1762$ ($0.1346$, $0.2207$) \\
& $\psi_{\gamma\gamma}$ & --- & 0.9345 (0.1917, NA)\tnote{b} \\
\hline
\hline
\multicolumn{4}{c}{\textbf{Coverage Probabilities}} \\
\hline
\hline
& \textbf{Para.} & \textbf{Reduced Model} & \textbf{Full Model} \\
\hline
& & Median (Range) & Median (Range) \\
\hline
\multirow{4}{*}{\textbf{\makecell[l]{Growth Factor \\ Means}}} 
& $\mu_{\eta_{0}}$ & $0.9485$ ($0.9300$, $0.9620$) & $0.9498$ ($0.9297$, $0.9688$) \\
& $\mu_{\eta_{1}}$ & $0.9445$ ($0.7560$, $0.9610$) & $0.9414$ ($0.8790$, $0.9700$) \\
& $\mu_{\eta_{2}}$ & $0.1560$ ($0.0040$, $0.5650$) & $0.1018$ ($0.0045$, $0.3819$) \\
& $\mu_{\gamma}$ & $0.9160$ ($0.3800$, $0.9490$) & $0.9464$ ($0.9210$, $0.9621$) \\
\hline
\multirow{4}{*}{\textbf{\makecell[l]{Growth Factor \\ Variances}}} 
& $\psi_{00}$ & $0.9315$ ($0.9050$, $0.9480$) & $0.9396$ ($0.9289$, $0.9631$) \\
& $\psi_{11}$ & $0.8140$ ($0.0030$, $0.9590$) & $0.9565$ ($0.9305$, $0.9672$) \\
& $\psi_{22}$ & $0.4715$ ($0.0010$, $0.9300$) & $0.8396$ ($0.6500$, $0.9354$) \\
& $\psi_{\gamma\gamma}$ & --- & $0.9690$ ($0.9340$, $0.9908$) \\
\hline
\hline
\end{tabular}
\label{tbl:Metric7}
\begin{tablenotes}
\small
\item[a] {--- indicates that the performance measures are not available from the reduced model.}
\item[b] {NA indicates that the bounds of performance measures are not available.}
\end{tablenotes}
\end{threeparttable}
\end{table}

\begin{figure}
\centering
  \includegraphics[width=1.0\linewidth]{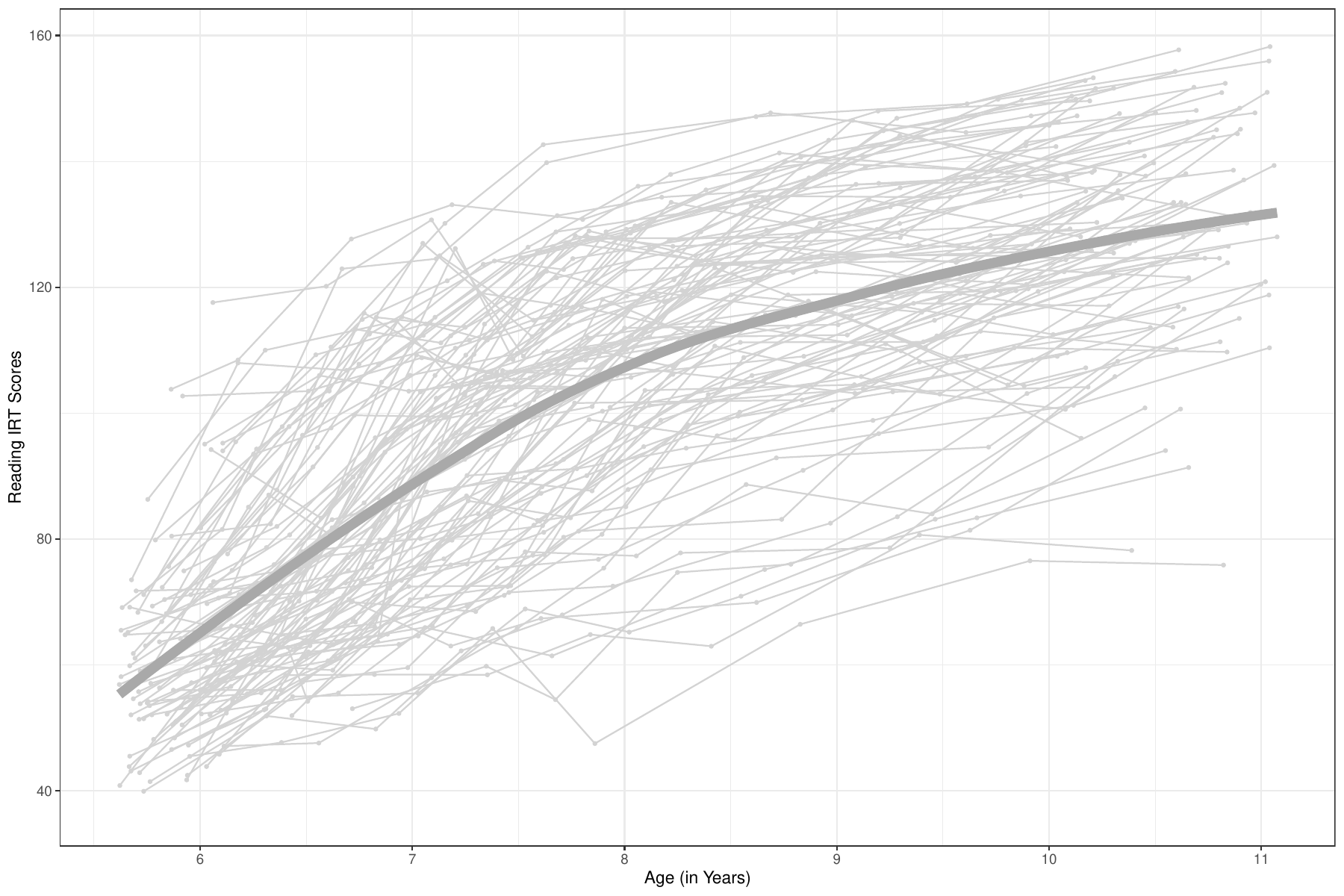}
  \caption{Observed Trajectories of the Development of Reading Ability and the Smooth Line\\
  Note: We randomly selected $100$ individuals from the $n=400$ subset of the ECLS-K:2011}
\label{fig:raw}
\end{figure}

\begin{figure}
\centering
  \includegraphics[width=1.0\linewidth]{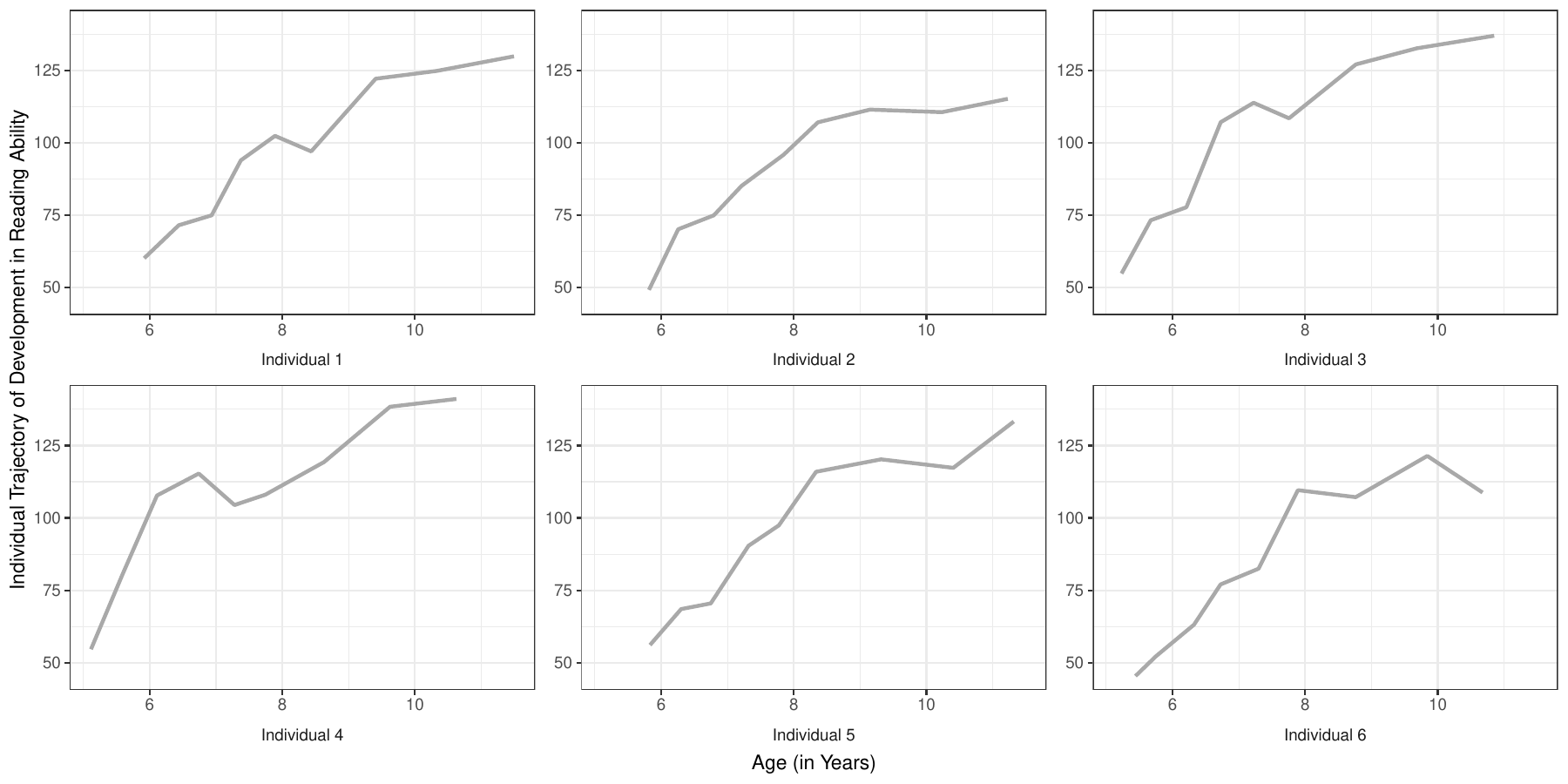}
  \caption{Individual Observed Trajectory of the Development of Reading Ability}
\label{fig:growth_ind}
\end{figure}

\end{document}